\documentclass[11pt,onecolumn,showpacs,preprintnumbers,amsmath,amssymb]{revtex4}

\usepackage{epsf}
\usepackage{graphicx}  
\usepackage{dcolumn}   
\usepackage{bm}        

\newcommand{\be}{\begin{equation}}
\newcommand{\en}{\end{equation}}
\newcommand{\bea}{\begin{eqnarray}}
\newcommand{\ena}{\end{eqnarray}}
\newcommand{\s}{\sqrt}

\newcommand{\lb}{\label}
\newcommand{\cp}{Chaplygin gas}
\begin{document}


\title{Crossing the phantom divide}

 \author{Hongsheng Zhang\footnote{Electronic address: hongsheng@kasi.re.kr} }
 \affiliation{ Shanghai United Center for
Astrophysics (SUCA), Shanghai Normal University, 100 Guilin Road,
Shanghai 200234,China}
  \affiliation{
 Korea Astronomy and Space Science Institute,
  Daejeon 305-348, Korea }
 \affiliation{Department of Astronomy, Beijing Normal University,
Beijing 100875, China}

\begin{abstract}
 The cosmic acceleration is one of the most significant cosmological discoveries
 over the last century. Following the more accurate data a more dramatic result
  appears:
  the recent analysis of the observation data (especially from SNe
  Ia)
  indicate that the time varying dark energy gives a better
  fit  than a cosmological constant, and in particular, the equation of state parameter
   $w$ (defined as the ratio of pressure to energy density) crosses $-1$ at some low redshift
   region. This crossing behavior is a serious challenge to
   fundamental physics. In this article, we review a number of approaches which try to explain this
   remarkable crossing behavior.  First we show
      the key observations which imply the crossing behavior. And then we
      concentrate on the theoretical progresses on the dark energy
      models which can realize the crossing $-1$ phenomenon. We
      discuss three kinds of dark energy models: 1. two-field
      models
      (quintom-like), 2. interacting models (dark energy interacts
      with dark matter), and 3. the models in frame of modified
      gravity theory (concentrating on brane world).

\end{abstract}

\pacs{95.36.+x  04.50.+h}

\maketitle

\newpage
\section{ The universe is accelerating}

    Cosmology is an old and young branch of science. Every nation had his
    own creative idea about this subject. However, till the 1920s about
    the unique observation which had cosmological significance was a dark sky
    at
    night. On the other hand, we did not prepare a proper
    theoretical foundation till the construction of general
    relativity. Einstein's 1917 paper is the starting point of
    modern cosmology \cite{e1}. The next
    mile stone was the discovery of cosmic expansion, that was the recession of galaxies and the
    recession velocity was proportional to the distance to us.

     Except some rare cases, our researches are always based on the
    cosmological principle, which says that the universe is
    homogeneous and isotropic. In the early time, this is only a
    supposition to simplify the discussions. Now we have enough
    evidences that the universe is homogeneous and isotropic at the
    scale larger than 100 Mpc. The cosmological principle requires
    that the metric of the universe is FRW metric,
   \bea
   ds^2=-dt^2+a^2(t)(dr^2+r^2d\Omega_2^2);
   \label{ping}
   \\
   ds^2=-dt^2+a^2(t)(dr^2+\sin(r)^2d\Omega_2^2);
   \label{qiu}
   \\
   ds^2=-dt^2+a^2(t)(dr^2+\sinh(r)^2d\Omega_2^2),
   \label{weiqiu}
   \ena
  depending on the spatial curvature, which can be Euclidean, spherical or
  pseudo-spherical. Here, $t$ is the cosmic time, $a$ denotes the
  scale factor, $r$ represents the comoving radial coordinate of the maximal
  symmetric 3-space, and $d\Omega_2^2$ stands for a 2-sphere. Which
  geometry serves our space is decided by observations. FRW metric
  describes  the kinetic evolution of the universe. To describe the
  dynamical evolution of the universe, that is, the function of $a(t)$, we need the gravity theory
  which ascribes the space geometry to matter. The present standard
  gravity theory is general relativity. In 1922 and 1924, Friedmann found that there
  was no static cosmological solution in general relativity, that is
  to say, the universe is either expanding or contracting \cite{f1}. To get a static
   universe, Einstein introduce the cosmological constant.
  However, even in the Einstein universe, where the contraction of
  the dust is exactly counteracted by the repulsion of the cosmological constant, the equilibrium is
  only tentative since it is a non-stationary equilibrium. Any small
  perturbation will cause it to contract or expand. Hence, in some
  sense we can say that general relativity predicts an expanding (or
  contracting) universe, which should be regarded as one of the most
  important prediction of relativity.

  In almost 70 years since the discovery of the cosmic expansion in 1929 \cite{h1}, people
    generally believe that the universe is expanding but the
    velocity is slowing down. People try to understand via observation that the
    universe will expand forever or become contracting at some
    stage. A striking result appeared in 1998, which demonstrated that the universe is accelerating rather than
    decelerating. Now we show how to conclude that our universe is
    accelerating. We introduce the standard general relativity,
    \be
    G_{\mu\nu}+\Lambda g_{\mu\nu}=8\pi G t_{\mu\nu},
    \label{ein}
    \en
    where $G_{\mu\nu}$ is Einstein tensor, $G$ is (4-dimensional)
    Newton constant, $t_{\mu\nu}$ denotes the energy-momentum
    tensor, $g_{\mu\nu}$ stands for the metric of a spacetime, $\mu, \nu$ run from 0 to 3. Throughout this article, we
    take a convention that $c=\hbar=1$  without special notation. Define a new energy momentum $T_{\mu\nu}$,
    \be
    8\pi G T_{\mu\nu}=8\pi G t_{\mu\nu} -\Lambda g_{\mu\nu}.
    \lb{newein}
    \en
   $T_{\mu\nu}$ has included the contribution of the cosmological
   constant, whose effect can not be distinguished with vacuum if we
   only consider gravity.

      The
    $00$ component of Einstein equation (\ref{ein}) is called
    Friedmann equation,
    \be
    H^2+\frac{k}{a^2}=\frac{8\pi G}{3} \rho,
    \label{fried}
    \en
   where $H=\frac{\dot{a}}{a}$ is the Hubble parameter, an overdot stands for the derivative with respect to the cosmic time,   $k$ is the
   spatial curvature of the FRW metric, for (\ref{ping}), $k=0$;
   for (\ref{qiu}), $k=1$; for (\ref{weiqiu}), $k=-1$, $\rho=-T^0_0$.
    Throughout this article, we take the signature
   $(-,+,...,+)$. The spatial component of Einstein equation can be replaced by the
   continuity equation, which is much more convenient,
   \be
   \dot{\rho}+3H(\rho+p)=0,
   \label{conti}
   \en
     where $p=T_1^1=T_2^2=T_3^3$. Here we use a supposition that the
     source of the universe $T_{\mu}^{\nu}$ is in perfect fluid
     form. Using (\ref{fried}) and (\ref{conti}), we derive the
     condition for acceleration,
    \be
    \frac{\ddot{a}}{a}=-\frac{4\pi G}{3} (\rho+3p).
    \en
   We see that the universe is accelerating when $\rho+3p<0$. We
   know that the galaxies and dark matter inhabit in the universe
   long ago. They are dust matters with zero pressure. Thus, if the
   universe is accelerating, there must exist an exotic matter with
   negative pressure or we should modify general relativity. The
   cosmological constant is a far simple candidate for this exotic
   matter, or dubbed dark energy. For convenience we separate the
   contribution of the cosmological constant (dark energy) from other sectors in
   the energy momentum, the Friedmann equation becomes,

   \be
   \frac{H^2}{H^2_0}=\Omega_{\Lambda 0}+\Omega_{m0}(1+z)^3+\Omega_{k0}(1+z)^2,
     \en
  where $z$ is the redshift, the subscript 0 denotes the present value of a quantity,
  \be \Omega_{\Lambda 0}=\frac{\Lambda}{8\pi H_0^2 G}, \label{omegal0} \en
  \be \Omega_{m0}=\frac{\rho_{m0}}{8\pi H_0^2 G}, \label{omegam0} \en
  \be \Omega_{k0}=-\frac{k}{ H_0^2 a_0^2}. \en

   The type Ia supernova is a most powerful tool to probe the
   expanding rate of the universe. In short, type Ia supernova is a
   supernova which just reaches the Chandrasekhar limit (1.4 solar mass) and
   then explodes. Hence they have the same local luminosity since they have roughly the same mass and
   the same exploding process. They are the standard candles in the unverse. We can get
   the distance of a type Ia supernova through its apparent magnitude. A sample of type Ia
   supernovae will generates a diagram of Hubble parameter versus distance, through which we
   get the information of the expanding velocity in the history of the universe.
     In 1998, two independent groups found
   that the universe is accelerating using the observation data of
   supernovae \cite{acce}. After that the data accumulate fairly quickly. The famous sample includes
   Gold04 \cite{gold04}, Gold06 \cite{gold06}, SNLS \cite{snls}, ESSENCE \cite{essence}, Davis07 \cite{davis07}, Union \cite{union}, Constitution \cite{constitution}.
   Here, we show some results of one of the most recent sample, Union \cite{union}, which is
   plotted by  $\chi^2$ statistics.

  \begin{figure}
\centering
 \includegraphics[totalheight=5in, angle=0]{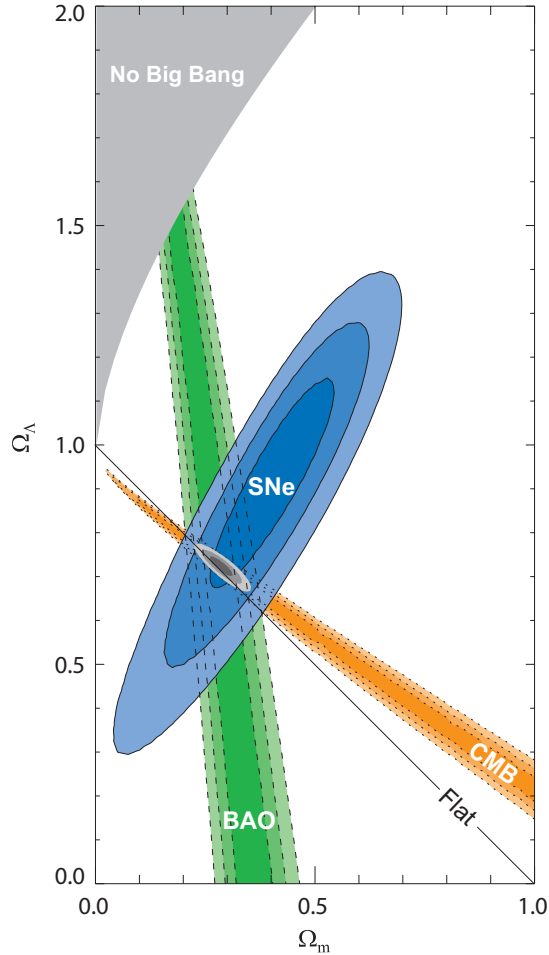}
\caption{The figure displays the counter constrained by SNe Ia (in
blue), by BAO (in green), by CMB (in yellow), and the joint
constraint by all the three kinds of observation (in black and
white). $\Omega_{\Lambda}$ is $\Omega_{\Lambda 0}$ in
(\ref{omegal0}), $\Omega_m$ is $\Omega_{m0}$ in (\ref{omegam0}).
This figure is borrowed from \cite{union}. }
 \label{lcdm}
 \end{figure}

 From fig \ref{lcdm}, we see that: 1. the universe is almost spatially flat, that is the curvature term $\Omega_{k0}$ is
 very small. 2. the present universe is dominated by cosmological
 constant (dark energy), whose partition is approximately 70\%, and
 the partition of dust is 30\%. We introduce a dimensionless
 parameter, the deceleration parameter $q$,
 \be
 q=-\frac{\ddot{a}a}{\dot{a}^2}.
 \en
 A negative deceleration parameter denotes acceleration.
 In a universe with dust and cosmological constant (which is called $\Lambda$CDM model), by definition
 \be
 q=\frac{1}{2}\Omega_{m}-\Omega_{\Lambda},
 \en
 whose present value $q_0=-0.55$. Hence the present universe is
 accelerating.

 The previous result depends on a special cosmological model,
 $\Lambda$CDM model in frame of general relativity. How about the conclusion if we only consider
 the kinetics of the universe?

 \begin{figure}
 \centering
  \includegraphics[totalheight=3in,angle=0]{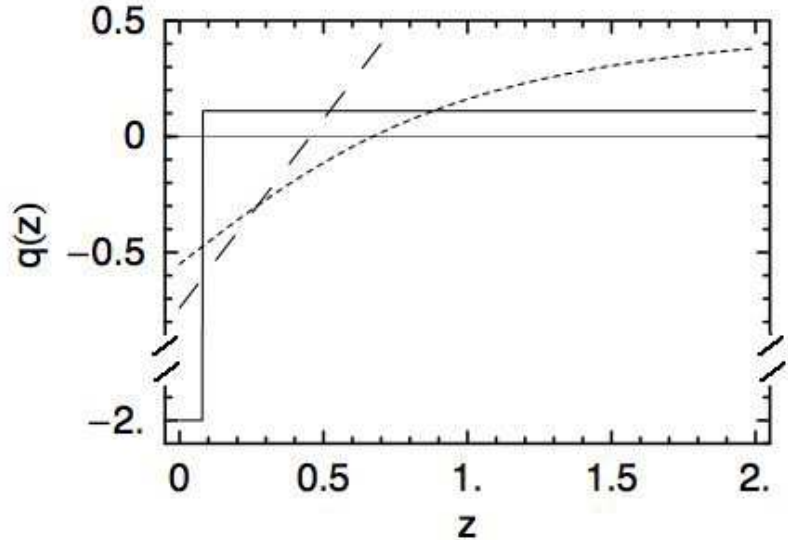}
  \caption{Kinetic universe vs dynamic universe. The fitting results
  of sudden transition model, linear expansion model, and
  $\Lambda$CDM model.  The solid line is the best-fit of the sudden transition model (the deceleration parameter jumps at some redshift);
  the long-dashed line denotes
  the best-fit of linear expansion model ($q=q_0+q_1z$, $q_1$ is a constant) \cite{gold04}; the short-dashed line represents the best-fit of $\Lambda$CDM
  model. From \cite{sha}.
  }
  \label{kinetics}
\end{figure}

  The simplest kinetic model is a sudden transition model, in which
  the deceleration parameter is a constant in some high redshift
  region and jumps to another constant at a critical redshift. The
  other simple choice is that the deceleration parameter is a linear
  function of $z$. We show the two kinetic models with the dynamic
  model $\Lambda$CDM in fig \ref{kinetics}. It is clear that the
  universe accelerates in the present epoch in all the three models.
  A more rigorous analyze shows that the evidence for an accelerating
  universe is fairly strong (more than 5 $\sigma$) \cite{sha}. So we
  should investigate it seriously.

  $\Lambda$CDM is the most simple model for the acceleration, which
  is a concordance model of several observations. As we shown in fig
  \ref{lcdm}, the counters of CMB, BAO, and SNe Ia have cross
  section, which almost laps over result of the joint fittings. However,
  $\Lambda$CDM has its own theoretical problems. Furthermore, it is
  found that a dynamical dark energy model fits the observation data
  better. Especially, there are some evidences that the equation of
  state (EOS) of dark energy may cross $-1$, which is a serious
  challenge to the foundation of theoretical physics.

  In the next section we shall study some problems of $\Lambda$CDM
  model and display that a dynamical dark energy model is favored by
  observations. We'll focus on the crossing behavior implied by the
  observation. In section III, we study 3 kinds of models with a
  crossing phantom divide dark energy. In section IV, we
  present the conclusion and more references of this topic.

 \section{A dark energy with crossing $-1$ EOS is slightly favored by observations}

  \subsection{ The problems of $\Lambda$CDM}
  $\Lambda$CDM has two famous theoretical problems.

  The first is the
  finetune problem. The effect of the vacuum energy can not be distinguished from the cosmological constant
  in gravity theory. We can calculate the vacuum energy by a
  well-constructed theory, quantum field theory (QFT), which says
  that the vacuum energy should be larger than the observed value by
  122 orders of magnitude, if QFT works well up to the Planck scale. In
  supersymmetric (SUSY) theory, the vacuum energy of the Bosons
  exactly counteracts the vacuum energy of Fermions, such that we obtain a zero vacuum energy.
  However, SUSY must break at the electro-weak scale. At that scale,
  the vacuum energy is still large than the observed value by 60
  orders of magnitude. So for getting a vacuum energy we observed,
  we should introduce a bare cosmological constant $\Lambda_{\rm
  bare}$. The effective vacuum energy $\rho_{\rm effect}$ then becomes,
  \be
  \rho_{\rm effect}=\frac{1}{8\pi G} \Lambda_{\rm bare}+\rho_{\rm vacuum}.
  \en
 $\frac{1}{8\pi G} \Lambda_{\rm bare}$ and $\rho_{\rm vacuum}$ have
 to almost counteract each other but do not exactly counteract each
 other, leaving a tiny tail which is smaller than the $\rho_{\rm vacuum}$ by 60
 orders of magnitude. Which mechanism can realize such a miraculous
 counteraction?

 The second problem is coincidence problem, which says that the
cosmological constant keeps a constant while the density of the dust
evolves as $(1+z)^3$ in the history of the universe, then why do
they approximately equal each other at ``our era"? Different from
the first problem, the second problem says the present ratio of dark
energy and dark matter is
 sensitively depends on the initial conditions.  Essentially, the coincidence problem is the problem of an
   unnatural initial condition. The densities of different species
   in  the universe redshift with different rate in the evolution of
   the universe, so if their densities coincidence in $our~era$,
   their density ratio must be a specific, tiny number in the
   $early~
   universe$. It is also a finetune problem, but a finetune problem of the initial condition.

   Except the above theoretical problems, $\Lambda$CDM also suffers from
   observation problem, especially when faced to the fine structure
   of the universe, including galaxies, clusters and voids.  Some specific observations differ
    from the predictions of $\Lambda$CDM (with standard partitions of dust and cosmological
    constant)
    at a level of 2$\sigma$ or higher. Six observations are summarized in \cite{problem lcdm}:
     1. scale velocity flows is much larger than the prediction of $\Lambda$CDM,
      2. Type Ia Supernovae (Sne Ia) at High Redshift are brighter
      than what $\Lambda$CDM indicates, 3. the void seems more empty than what $\Lambda$CDM predicts,
      4. the cluster haloes look denser than  what $\Lambda$CDM
      says,
        5. the density function of galaxy haloes is smooth, while $\Lambda$CDM indicates a cusp in the
        core,
   6. there are too much disk galaxies than  the prediction of
   $\Lambda$CDM. We do not fully understand the dynamics
   and galaxies and galaxy clusters, that is, the gravitational perturbation theory at the small scale. The agreement may
   approve when we advance our perturbation theory with a cosmological constant and the simulation methods at the small scale.
   However, in the cosmological scale, there are also some evidences
   that the dark energy is dynamical, including no. 2 of the previous
   6 problems.

   \subsection{crossing $-1$}
\begin{figure}
 \centering
  \includegraphics[totalheight=2.3in,angle=0]{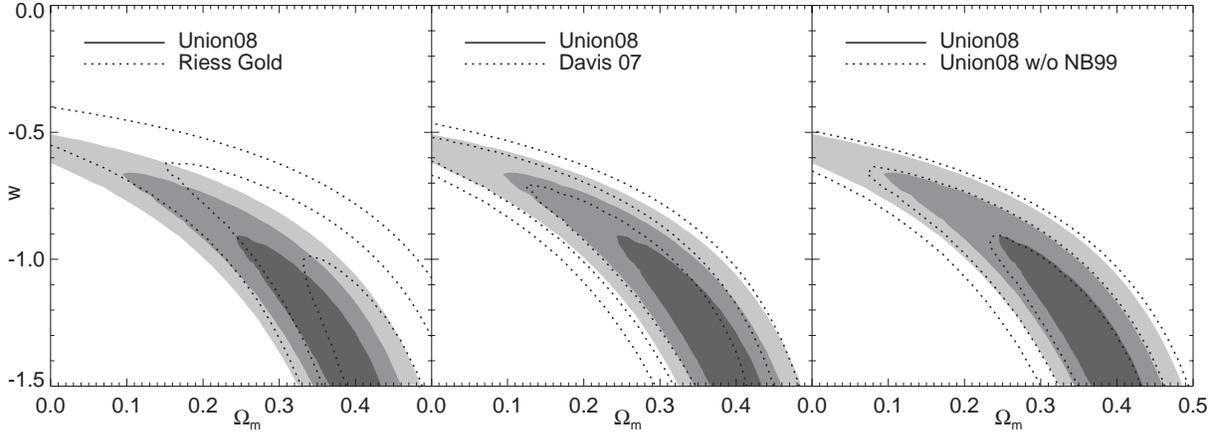}
  \caption{The EOS of dark energy fitted by SNe Ia in a spatially flat universe, the contours display  68.3~\%, 95.4~\% and 99.7\% confidence level on
 $w$ and $\Omega_{m0}$ ($\Omega_{m}$ in the figure). The results of the
 Union set are  shown as filled contours. The empty contours, from left to right,
 show the results of the Gold sample, Davis 07, and the Union without SCP nearby data. From \cite{union}.
  }
  \label{eos3}
\end{figure}

   With data accumulation, observations which favor dynamical dark
   energy become more and accurate. Now we loose the condition that
   $p=-\rho$ for the exotic matter (dark energy) which accelerates the universe.
   We go beyond the $\Lambda$CDM model. We permit that the EOS of dark
   energy is not exactly equal to $-1$, but still a constant. The
   fitting results by different samples of SNe Ia are displayed in
   fig \ref{eos3}. We see that although a cosmological constant is
   permitted, the dark energy whose EOS  $<-1$ is favored by SNe
   Ia. The essence whose EOS is less than $-1$ is called phantom,
   which can be realized by a scalar field with negative kinetic
   term. The action for phantom $\psi$ is
   \be
   S_{\rm ph}=\int d^4x \sqrt{-g} \left(\frac{1}{2}\partial_{\mu} \psi
   \partial^{\mu} \psi-U(\psi)\right),
   \label{actionphantom}
   \en
   where $\sqrt{-g}$ is the determinate of the metric, the lowercase
   Greeks run from 0 to 3, $U(\psi)$ denotes the potential of the
   phantom. In an FRW universe, the density and pressure
   of the phantom (\ref{actionphantom}) reduce to
   \be
   \rho=-\frac{1}{2}\dot{\psi}^2+U,
   \lb{rhophan}
   \en
   \be
   p=-\frac{1}{2}\dot{\psi}^2-U,
   \lb{pphan}
   \en
  respectively. Now the EOS of the phantom  $w$ is given by,
  \be
  w=\frac{p}{\rho}=\frac{-\frac{1}{2}\dot{\psi}^2-U}{-\frac{1}{2}\dot{\psi}^2+U},
  \en
  which is always less than $-1$ for a positive $U$. It seems that
  phantom is proper candidate for the dark energy whose EOS less
  than $-1$ \cite{phantom}. It is very famous that a phantom field
  is unstable when quantized since the energy has no lower bound. It will transit
  to  a lower and lower energy state. In this article we have no time to discuss
  this important topic for phantom dark energy. We would point out the basic idea for this issue, which
  requires the life time of the phantom is much longer than the age of the universe such that
  we still have no chance to observe the decay of the phantom, though it is fundamentally unstable.  For references, see
  \cite{stablephan}. A completely regular quantum stress with $w <
  -1$ is suggested in \cite{phanquan}.

 If the dark energy really behaves as phantom at some low redshift
  region, it is  an unusual discovery. But the dark energy may be
  more fantastic. In some model-independent fittings, the EOS of
  dark energy crosses $-1$, which is a really remarkable property
  and  a serious challenge to our present theory of fundamental
  physics.

\begin{figure}
 \centering
  \includegraphics[totalheight=2.6in,angle=0]{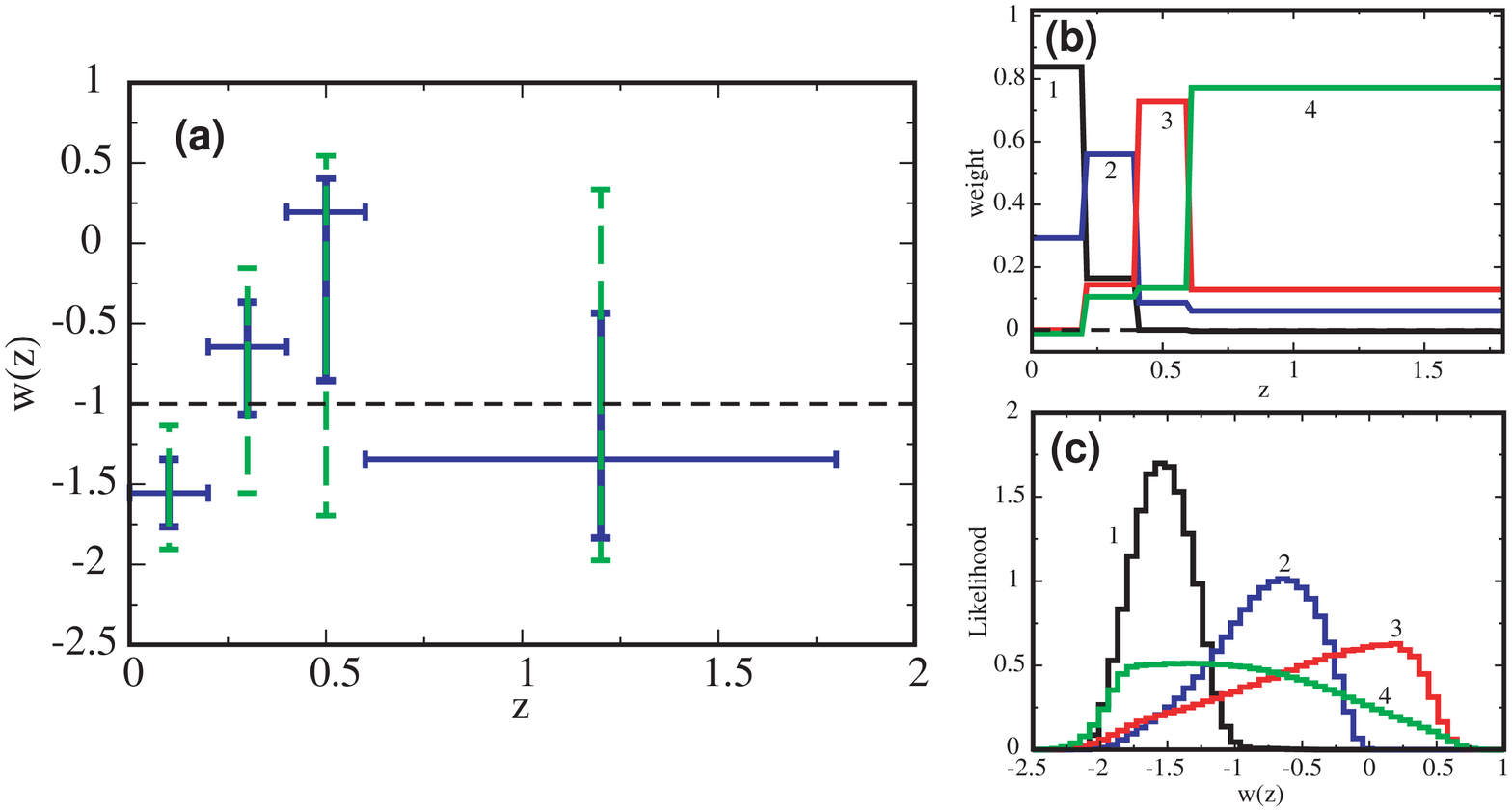}
  \caption{Panel (a): Uncorrelated band-power estimates of the EOS
 $w(z)$ of dark energy by SNe Ia (Gold set \cite{gold04}). Vertical error bars
 show the 1 and 2-$\sigma$ error bars (in blue and green, respectively). The horizontal error bars denote the
 data bins used in \cite{cross1}.   Panel (b): The window
 functions for each bin from low redshift to high redshift. Panel (c):
the likelihoods of $w(z)$
 in the bins from low redshift to high redshift.
 From \cite{cross1}.
  }
  \label{cross1f}
\end{figure}

  Pioneer results of the crossing $-1$ of EOS of dark energy appeared in \cite{cross0, cross1}.
   Fig \ref{cross1f} illuminates that the EOS of dark energy may cross $-1$ in some
  low redshift. In fig \ref{cross1f}, the Gold04 data are applied, a uniform prior of $0.22\leq
\Omega_{m0} \leq 0.38$ is assumed, and a spatially flat universe is
the working frame.

\begin{figure}
 \centering
  \includegraphics[totalheight=3.3in,angle=0]{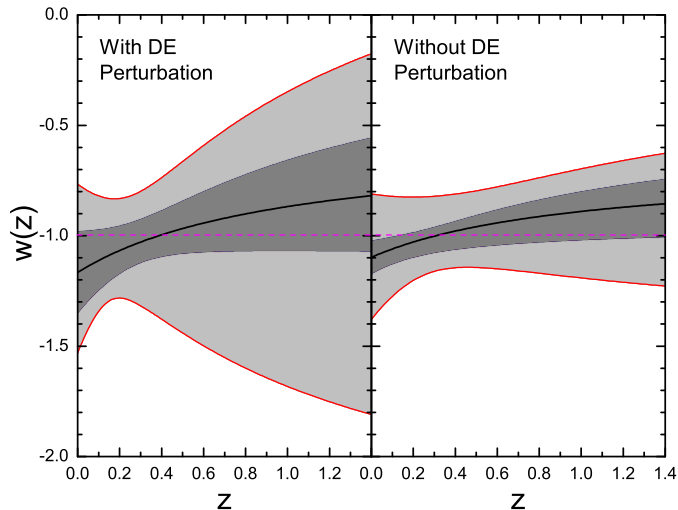}
  \caption{Constraints on the EOS of w(z) by WMAP3 \cite{wmap3} and Gold04
  \cite{gold04}. The light grey region denotes 2
  $\sigma$ constraint,
  while the dark grey for 1$\sigma$ constraint. The left panel shows
  the constraint with dark energy perturbation, while the right
  displays the result without dark energy perturbation.
 From \cite{cross2}.
  }
  \label{cross2f}
\end{figure}

  The perturbation of the dark energy will growth if its EOS is not
  exactly $-1$ in the evolution history of the universe.
   Hence to  fit a model with dynamical dark energy with observation, the perturbation of the
  dark matter should be considered in principle. Such a study was
  presented in \cite{cross2}, in which a parametrization of the EOS of the dark energy with two constant $w_0,~w_1$ was applied,
  \be
  w=w_0+w_1\frac{z}{1+z}.
    \en
    The result is shown in fig
  \ref{cross2f}.
 We see from fig \ref{cross2f} that there is a mild tendency that
 the EOS of the dark energy cross $-1$. For a more general parametrization of EOS for dark
 energy, see \cite{zhupara}.

 With more and accurate data,
 the possibility of crossing $-1$ (phantom divide) seems a little
 more specific, see for example \cite{crossmany}. This crossing
 behavior is a significant challenge for theoretical physics. It was
 proved that the EOS of dark energy can not cross the phantom divide
 if 1. a dark energy component with an arbitrary scalar-field Lagrangian,
  which has a general dependence on the field itself and its first derivatives,
  2. general relativity holds and 3. the spatially flat Friedmann
  universe \cite{nogo}, for a more detailed proof, wee the appendix of \cite{nogo2}. Thus realizing such a crossing is not a trivial work. In
  the next section we investigate the theoretical progresses for
  this extraordinary phenomenon.

  \section{three roads to cross the phantom divide}
 To cross the phantom divide, we must break at least one of the
 conditions in \cite{nogo}. Now that the dark energy behaves as quintessence at some stage
 , while evolves as phantom at the other stage, a natural suggestion is that we should
 consider a 2-field model, a quintessence and a phantom. The
 potential is carefully chosen such that the quintessence dominates the
 universe at some stage while the phantom dominates the universe at the other stage. It was invented
 a name for such 2-field model, ``quintom" . There are also some
 varieties of quintom, such as hessence. We introduce these 2-field
 models in the first subsection. The next road is to consider an
 interacting model, in which the dark energy interacts with dark
 matter. The interaction can realize the crossing behavior which is
 difficult for independent dark energy. We shall
 study the interacting models in subsection B. The other possibility
 is that general relativity fails at the cosmological scale. The
 ordinary dark energy candidates, such as quintessence or phantom,
 can cross the phantom divide in a modified gravity theory. We
 investigate this approach in subsection C.

 \subsection{2-field model}
  A typical 2-field model is the quintom model, which was proposed
  in \cite{quintom1}, and was widely investigated later
  \cite{quintom2}. Generally
  , the action  of a universe with quintom dark energy $S$ is
  \be
  S=\int d^4x \sqrt{-g}\left(\frac{R}{16\pi G}+{\cal L_{\rm stuff}} \right)
  ,
  \en
  where $R$ is the Ricci scalar, $\cal L_{\rm stuff}$ encloses all
  kinds of the stuff in the universe, for instance the dust matter,
  radiation, and quintom. At the late universe, the radiation can be
  negligible. So, often we only consider the dark energy, here
  quintom $\cal L_{\rm quintom}$, and dust matter $\cal L_{\rm dm}$,
  \be
  {\cal L_{\rm quintom}}=-\frac{1}{2}\partial_{\mu} \phi
   \partial^{\mu} \phi+\frac{1}{2}\partial_{\mu} \psi
   \partial^{\mu} \psi-W(\phi,~\psi).
   \label{quin}
   \en
  In (\ref{quin}) the first term is the kinetic term of an ordinary
  scalar, the second term is the kinetic term of a phantom, and
  $W(\phi,~\psi)$ is an arbitrary function of $\phi$ and $\psi$.  In an FRW
 universe,
 the density and pressure of the quintom are
 \be
   \rho=\frac{1}{2}\dot{\phi}^2-\frac{1}{2}\dot{\psi}^2+W,
   \en
   \be
   p=\frac{1}{2}\dot{\phi}^2-\frac{1}{2}\dot{\psi}^2-W.
   \en
  Hence, the EOS of the quintom $w$ is
  \be
  w=\frac{\frac{1}{2}\dot{\phi}^2-\frac{1}{2}\dot{\psi}^2-W}{\frac{1}{2}\dot{\phi}^2-\frac{1}{2}\dot{\psi}^2+W}.
  \en
 $w=-1$ requires
 \be
 \dot{\phi}^2=\dot{\psi}^2.
 \en
  We see that in a quintom model, we do not require a static field (a field with zero kinetic term or a field at ground state) to
  get a cosmological constant. We only need that $\psi$ and $\phi$
  evolves in the same step. $w<-1$ implies,
  \be
  \dot{\phi}^2-\dot{\psi}^2<0,
  \en
  if
  \be
  \frac{1}{2}\dot{\phi}^2-\frac{1}{2}\dot{\psi}^2+W>0;
  \label{rhocon1}
  \en
  and
  \be
  \dot{\phi}^2-\dot{\psi}^2>0,
  \en
  if
 \be
  \frac{1}{2}\dot{\phi}^2-\frac{1}{2}\dot{\psi}^2+W<0.
  \label{rhocon2}
  \en
  (\ref{rhocon2}) yields an unnatural  physical result,that is, the density
  of dark energy is negative. However, this is not as serious as the
  first glance, since we have little knowledge of the dark energy
  besides its effect of gravitation. Several evidences imply that we
  should go beyond the standard model of the particle physics when
  we describe dark energy. There are a few dark energy models
  permit density of the dark energy, or a component of it is negative (at the same time keep
  the total density positive), for example, see \cite{negden1, negden2}. But,
  for a model with only two components, a dust and a quintom, it is
  difficult to set a negative density dark energy. In that case we
  need too much dust than we observed or a big curvature term. In
  the following text of this section, we only consider a dark energy with
  positive density. So $w>-1$ implies,
 \be
  \dot{\phi}^2-\dot{\psi}^2>0.
  \en
  In summary, if the kinetic term of the quintessence dominates that
  of phantom, the quintom behaves as quintessence; else it behaves
  as phantom. We should select a proper potential to make
  quintessence and phantom dominate alternatively such that we can
  realize the crossing behavior.

  A simple choice of the potential is that the quintessence and the
  phantom do not interact with each other, which requires,
  $W(\phi,\psi)=V(\phi)+U(\psi)$. The exponential potential is an important
  example which can be solved exactly in the quintessence model (a toy universe only composed by quintessence). In
 addition, we know that such exponential potentials of scalar fields
  occur naturally in some fundamental theories such as string/M
  theories. We introduce  a model with such potentials in
  \cite{quintomguo}, in which the potential $V(\phi,\psi)$ is given by
  \be
  W(\phi,\psi)=V(\phi)+U(\psi)=A_\phi e^{-\lambda_\phi \kappa \phi}+A_\psi
 e^{-\lambda_\psi \kappa \psi},
  \en
 where $A_\phi$ and $A_\psi$ are the amplitude of the potentials,
 $\kappa^2=8\pi G$, $\lambda_\phi$ and $\lambda_\psi$ are two
 constants. Since there is no direct couple between the quintessence and the phantom, the equations
 of motion of the quintessence and the phantom are two independent
 equations,
 \be
 \ddot{\phi}+3H\dot{\phi}+\frac{dV}{d\phi} = 0,
 \label{eomphi}
 \en

 \be
 \ddot{\psi}+3H\dot{\psi}-\frac{dU}{d\psi} = 0.
 \label{eompsi}
 \en
 The continuity equation of the dust reads,
 \be
 \rho_{\rm dust}+3H\rho_{\rm dust}=0,
 \label{contidust}
 \en
 where $\rho_{\rm dust}$ denotes the density of the dust.
 The method of dynamical system has been widely used in cosmology.
 This method can offer a clear history of the cosmic evolution,
 especially the final states of the university. For applying this
 method, first we define the following dimensionless variables,
\begin{eqnarray}
x_\phi \equiv \frac{\kappa \dot{\phi }}{\sqrt{6}H}&,& \quad y_\phi
 \equiv \frac{\kappa \sqrt{V_\phi}}{\sqrt{3}H}, \nonumber \\
x_\psi \equiv \frac{\kappa \dot{\psi _i}}{\sqrt{6}H}&,& \quad y_\psi
 \equiv \frac{\kappa \sqrt{V_\psi}}{\sqrt{3}H}, \\
z \equiv \frac{\kappa \sqrt{\rho_{\rm dust}}}{\sqrt{3}H}&,&
\nonumber
\end{eqnarray}
the evolution equations (\ref{eomphi})-(\ref{contidust}) become,
  \bea
\label{AS1} x'_\phi &=&
-3x_\phi\left(1+x_{\phi}^2-x_{\psi}^2-\frac{1}{2}
 z^2\right)+\lambda_\phi \frac{\sqrt{6}}{2}y_{\phi}^2\,, \\
y'_\phi &=& 3y_\phi\left(-x_{\phi}^2+x_{\psi}^2+\frac{1}{2}
 z^2-\lambda_\phi \frac{\sqrt{6}}{6}x_\phi\right), \\
x'_\psi&=&-3x_\psi\left(1+x_{\phi}^2-x_{\psi}^2-\frac{1}{2}
 z^2\right)-\lambda_\psi \frac{\sqrt{6}}{2}y_{\psi}^2\,, \\
y'_\psi &=& 3y_\psi\left(-x_{\phi}^2+x_{\psi}^2+\frac{1}{2}
 z^2-\lambda_\psi \frac{\sqrt{6}}{6}x_\psi\right), \\
z' &=& 3z\left(-x_{\phi}^2+x_{\psi}^2+\frac{1}{2}
 z^2-\frac{1}{2}\right),
\label{AS2}
 \ena
 in which a prime denotes derivative with respect to $\ln a$. Generally, $z$ in the above set will not be confused with redshift.
   The five equations in this
 system are not independent. They are constrained by Fridemann
 equation,
\begin{equation}
\label{FE} H^2=\frac{\kappa ^2}{3}\left(\frac{1}{2}\dot{\phi}^2+V
 -\frac{1}{2}\dot{\psi}^2+U+\rho_{\rm dust} \right),
\end{equation}
which becomes
 \be
 x_{\phi}^2+y_\phi^2-x_{\psi}^2+y_\psi^2+z^2=1.
 \en
 with the dimensionless variables defined before. The critical
 points dwell at $x_\phi'=y_\phi'=x_\psi'=y_\psi'=z'=0$. We present
 the result in table \ref{tabcri}.
 \begin{table}
\begin{tabular}{c c c c c c c} \hline
Label & $x_\psi$ & $y_\psi$ & $x_\phi$ & $y_\phi$ & z
 & Stability \\ \hline
$K$ & $-x_{\psi}^2+x_{\phi}^2=1$ & 0 & & 0 & 0 & unstable \\
$P$ & $-\frac{\lambda_\psi}{\sqrt{6}}$
 & $\sqrt{ (1+\frac{\lambda_\psi^2}{6})}$ & 0 & 0 & 0 & stable  \\
$S$ & 0 & 0 & $\frac{\lambda_\phi}{\sqrt{6}}$ &
 $\sqrt{ (1-\frac{\lambda_\phi^2}{6})}$ & 0 & unstable  \\
$F$ & 0 & 0 & 0 & 0 & 1 & unstable \\
$T$ & 0 & 0 & $\frac{3}{\sqrt{6}\lambda_\phi}$
 & $\frac{\sqrt{{3}}}{\lambda_\phi}$
 & $\sqrt{1-\frac{3}{\lambda_\phi^2}}$ & unstable \\ \hline
\end{tabular}
\caption{ The critical points, from \cite{quintomguo}}
 \label{tabcri}
 \end{table}
 For detailed discussion of the critical points, see
 \cite{quintomguo}. We would like to show a numerical example in which the EOS of the quintom
 crosses the phantom divide. Fig 6 illuminates that the
 EOS crosses $-1$.

  \begin{figure}
\begin{center}
\includegraphics[scale=1.]{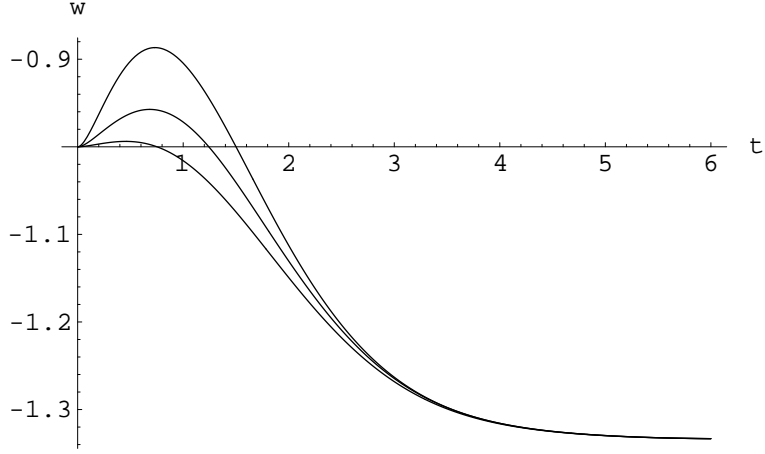}
\caption{The evolution of the effective equation of state of the
phantom and normal scalar fields with $W(\phi,\sigma)$ for the case
$\lambda_{\phi}=1$. From \cite{quintomguo}.}
\end{center}
\lb{fcrossguo2}
\end{figure}

   The previous quintom model includes two fields, which are completely independent and rather
   arbitrary. We can impose some symmetry in the quintom model. An
   interesting model with an internal symmetry between the two fields which work as dark energy is hessence \cite{hessence}.
  Rather than two uncorrelated fields, we consider one complex scalar field with internal symmetry between the real and the imaginary parts,
 \be
 \lb{daphi}
 \Phi=\phi_1+i\phi_2,
 \en
 with a Lagrangian density
 \be
 \lb{daphiL}
 {\cal L}_{\rm hess}=-\frac{1}{4}\left[(\partial_\mu
 \Phi)^2+(\partial_\mu \Phi^*)^2\right] -V(\xi,\Phi^\ast)=-\frac{1}{2}\left[\,(\partial_\mu \xi)^2-\xi^2 (\partial_\mu
\theta)^2\,\right]-V(\xi),
 \en
 which is invariant under the transformation,
 \bea
 \phi_1\to\phi_1\cos\alpha-i\phi_2\sin\alpha,\\
 \phi_2\to-i\phi_1\sin\alpha+\phi_2\cos\alpha,
 \lb{internaltran}
 \ena
 if the potential is only a function of $\Phi^2+(\Phi^*)^2$.
 For convenience, in (\ref{daphiL}) we have introduced two new variables $(\xi,\theta)$,
 \be
 \phi_1=\xi\cosh\theta,~~~~~~~\phi_2=\xi\sinh\theta,
 \en
 which are defined by
 \be
 \xi^2=\phi_{1}^2-\phi_{2}^2,~~~~~~~\coth\theta=\frac{\phi_1}{\phi_2}.
 \en
 The equations of motion of $\xi$ and $\theta$ are
\be
 \ddot{\xi}+3H\dot{\xi}+\xi\dot{\theta}^2+\frac{dV}{d\xi}
 =0,
 \lb{eomh1}
 \en
 \be
  \xi^2
 \ddot{\theta}+(2\xi\dot{\xi}+3H\xi^2)\dot{\theta}=0.
 \lb{eomh2}
 \en
  Clearly, $\xi$ and $\theta$ couple to each other.
 The
 pressure and density of the hessence read,
 \be
 p_{\rm hess}=\frac{1}{2}\left(\dot{\xi}^2-\xi^2
\dot{\theta}^2\right)-V(\xi),
 \lb{ph}
 \en
 \be
\rho_{\rm hess}=\frac{1}{2}\left(\dot{\xi}^2-\xi^2
\dot{\theta}^2\right)+V(\xi),
 \lb{rhoh}
 \en respectively.
  The EOS of hessence, playing as dark energy,
 \be
 w=\frac{\frac{1}{2}\left(\dot{\xi}^2-\xi^2
 \dot{\theta}^2\right)-V(\xi)} {\frac{1}{2}\left(\dot{\xi}^2-\xi^2
 \dot{\theta}^2\right)+V(\xi)}.
 \en
  Qualitatively, hessence evolves as quintessence  when $\dot{\xi}^2\geq\xi^2 \dot{\theta}^2$, while as phantom when
   $\dot{\xi}^2<\xi^2 \dot{\theta}^2$. The Lagrangian (\ref{daphiL})
   does not include $\theta$, hence the canonical momentum
   $\pi_{\theta}^{\mu}$
   corresponding to the cyclic coordinate $\theta$ are  conserved
   quantities,
   \be
   \pi_{\theta}^{\mu}=\frac{\partial ({\cal L}_{\rm hess}\s{-g})}{\partial(\partial_\mu
   \theta)}.
   \en
   In an FRW universe, only $\pi_\theta^0$ exists. We define a
   conserved quantity $Q$ which is proportional to $\pi_\theta^0$,
   \be
   Q=a^3 \xi^2\dot{\theta}.
   \en
   With this conserved quantity, the EOS becomes,
   \be
   w=\frac{\frac{1}{2}\dot{\xi}^2-\frac{Q^2}{2a^6 \xi^2}-V(\xi)}{
\frac{1}{2}\dot{\xi}^2-\frac{Q^2}{2a^6 \xi^2}+V(\xi)},
    \en
    which is only a function of $\xi$.
   The Friedmann equations
 read as
 \bea
  H^2=\frac{8\pi
 G}{3}\left[\rho_{\rm dust}+\frac{1}{2}\left(\dot{\xi}^2-\xi^2
 \dot{\theta}^2\right)
 +V(\xi)\right],
 \ena
  where
 $\rho_{\rm dust}$ is the energy density of dust. The continuity
 equation of dust is (\ref{contidust}). The continuity equations of
 hessence are identical to the equations of motion (\ref{eomh1})
 and (\ref{eomh2}). Then the system is closed and we present a
 numerical example in fig 7. Evidently, the EOS of hessence, playing the role of dark energy,
 crosses $-1$ at about $a=0.95$ ($z=0.06$).

  \begin{figure}
\begin{center}
\includegraphics[scale=1.]{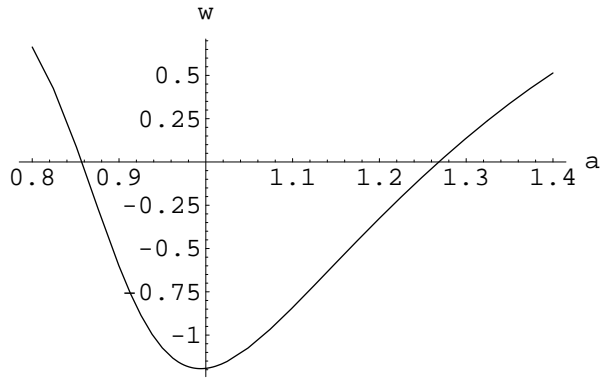}
\caption{The EOS of hessence $w$ as a function of scale factor with
the potential $V(\xi)=\lambda\xi^4$. The parameters for this plot
are as follows: $\Omega_{m0}=\rho_{m0}/(3H_0^2)=0.3$, $\lambda=5.0$,
 $Q=1.0$, $a_0=1$ and the unit $8\pi G=1$. From \cite{hessence}.
}
\end{center}
\lb{fighessence}
\end{figure}

  After the presentation of hessence model, several aspects of this model have been investigated, including to avoid the big rip
  \cite{wei2}, attractor solutions for general hessence \cite{hess3},
  reconstruction of hessence by recent observations \cite{hess4}, dynamics of hessence in frame of loop quantum
  cosmology \cite{hess5}, and holographic hessence
  model\cite{hess6}.

 \subsection{interacting model}
  Two-field model is a natural and obvious construction to realize
  the crossing $-1$ behavior of dark energy. However, there are two
  many parameters in the set-up, though we can impose some
  symmetries to reduce the parameters to a smaller region. One
  symmetry decrease one parameter, but we have little clue to impose
  the symmetries since we have no evidence in the ground labs.

  Interaction is a universal phenomenon in the physics world. An
  interaction term is helpful to cross the phantom divide. To
  illuminate this point, we first carefully analyze the previous
  observations which imply the crossing. Both the results of
  \cite{cross1} and \cite{cross2}, which are shown in fig \ref{cross1f} and fig \ref{cross2f} respectively,
  are derived with a presupposition, that is, the dark energy
  evolves freely. In fact, what we observed is the effective EOS of
  the dark energy in the sense of gravity at the cosmological scale. When we suppose it evolves freely, we find that
  its EOS may cross the phantom divide. We can demonstrate for an
  essence with (local) EOS$<-1$, the cosmological effective EOS can cross $-1$ by aids of an interacting term.
  For the case with interaction, the continuity equation for dark
  energy becomes,
  \be
  \dot{\rho}_{\rm de}+3H(\rho_{\rm de}+p_{\rm de})=-\Gamma,
 \lb{contide}
  \en
 or
  \be
  \dot{\rho}_{\rm de}+3H(\rho_{\rm de}+p_{\rm
  de}+\frac{\Gamma}{3H})=0.
 \lb{contide}
  \en
 Here $\rho_{\rm de}$ is the density of dark energy, $p_{\rm de}$
 denotes the local pressure measured in the lab (if we can measure),
 $\Gamma$ stands for the interaction term, and $p_{\rm eff}=p_{\rm
  de}+\frac{\Gamma}{3H}$ is the effective pressure in the
  cosmological sense. In a universe without expanding or
  contracting, $H=0$, the interaction does no effect on the
  continuity equation, or energy conservation law, and thus does not
  yield surplus pressure \footnote{One may think that $\Gamma/H$ is meaningless when $H=0$. But in fact, in most realistic cases, we always
  assume that $\Gamma$ is proportional to $H$.}.  Two special cases are interesting:
  1. $\frac{\Gamma}{3H}$ is a constant, under which the interaction term
  contributes a constant pressure throughout the history of the universe. 2. $\frac{\Gamma}{3H\rho_{de}}$ is a constant,
  under which the interaction term contributes a constant EOS in the history of the universe.
  In frame of a quintessence or phantom dark energy, the interaction
  term $\frac{\Gamma}{3H\rho_{de}}$ only shifts the EOS up or down
  by a constant distance in the $w-z$ plane, without changing the
  profile of the curve of $w$. While the term $\frac{\Gamma}{3H}$
  shifts the pressure, which can change the EOS significantly since
  the density $\rho_{\rm de}$ is a variable  in the history of the
  universe.

 If the dark energy can couple to some stuff of the universe, the
 dark matter is the best candidate.  Although non-minimal coupling between the dark energy and ordinary
matter fluids  is strongly restricted by the  experimental tests in
the solar system \cite{will}, due to the unknown nature of the dark
matter as part of the background, it is possible to have
non-gravitational interactions between the dark energy and the dark
matter components, without conflict with the experimental data.
 The continuity equation for dust-like dark matter reads,
 \be
 \dot{\rho}_{\rm dm}+3H\rho_{\rm dm}=\Gamma.
 \lb{contidm}
  \en

  Based on the previous discussion, we assume a most simple case
 \be
 \Gamma=H\delta \rho_{\rm dm},
 \en
 where $\delta$ is constant \cite{guointer, weiinter, amen}. This interaction term shifts a constant
 to the EOS of $dark
 ~matter$, that is, it is no longer evolving as $(1+z)^{3}$.  We
 uniformly deal with quintessence and phantom, which are often labeled by
 $X$, with a constant EOS $w_X$.  So, the continuity equation of dark
 energy can be written as,
 \be
  \dot{\rho}_{\rm X}+3H(\rho_{\rm X}+w_X \rho_{\rm X})=-H\delta \rho_{\rm
  dm}.
 \lb{contiX}
  \en
  Integrating (\ref{contidm}), we derive
  \be
 \label{rhodm}
 \rho_{\rm dm}=\rho_{\rm dm0}a^{-3+\delta} =\rho_{\rm dm0}
 (1+z)^{3-\delta}.
  \en
 Substituting to (\ref{contiX}), we reach
 \be
 \label{rhoX}
\rho_X =\rho_{X0} (1+z)^{3(1+w_X)}  +\rho_{dm0}
\frac{\delta}{\delta+3w_X} \left[
(1+z)^{3(1+w_X)}-(1+z)^{3-\delta}\right].
 \en
 Only from the above equation, we can extract the effective EOS of the dark energy.
 To see this point, we make a short discussion. In a dynamical universe with interaction, the
  effective EOS of dark energy reads,
  \be
  w_{de}=\frac{p_{\rm eff}}{\rho_{\rm de}}=\frac{p_{\rm de}+\Gamma/3H}{\rho_{\rm de}}=-1+\frac{1}{3}\frac{d \ln \rho_{de}}{d \ln
  (1+z)}.
     \en
   Clearly, if $\frac{d \ln \rho_{de}}{d \ln
  (1+z)}$ is greater than 0, dark energy evolves as quintessence; if $\frac{d \ln \rho_{de}}{d \ln
  (1+z)}$ is less than 0, it evolves as phantom; if $\frac{d \ln \rho_{de}}{d \ln
  (1+z)}$ equals 0, it is just cosmological constant. In a more
  intuitionistic way, if $\rho_{de}$ decreases and then increases
  with respect to redshift (or time), or increases and then
  decreases, which implies that EOS of dark energy crosses phantom
  divide. So, some time we directly use the evolution of density of dark energy
  to describe the EOS of it. There is a more important motivation to
  use the density directly: the density is
 more closely related to observables, hence is more tightly
 constrained for the same number of redshift bins used
 \cite{wangyun}.

 The derivative of $\rho_X$ with respect to (1+z) reads,
 \be
 \frac{d\rho_X}{d(1+z)}=3(1+w_X)\rho_{X0} (1+z)^{2+3w_X}  +\rho_{dm0}
\frac{\delta}{\delta+3w_X} \left[
 3(1+w_X)(1+z)^{2+3w_X}-(3-\delta)(1+z)^{2-\delta}\right].
 \en
 If $\frac{d\rho_X}{d(1+z)}=0$ at some redshift $z=z_c$, the
 effective EOS crosses $-1$. The result is illuminated by fig 8, in which we set $z_c=0.3$ as
 an example. This figure displays the corresponding $w_X$ when
 one fixes a $\delta$, or vice versa if we require the EOS crosses
$-1$ at
 $z_c=0.3$. This is an original figure plotted for this review
 article.

 \begin{figure}
\begin{center}
\includegraphics[scale=1.2]{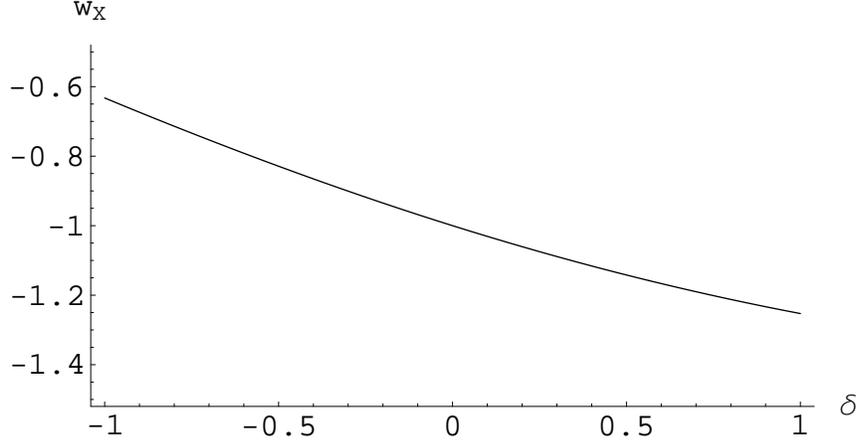}
\caption{ $w_X$ vs $\delta$ under the condition
$\frac{d\rho_X}{d(1+z)}=0$.}
\end{center}
\lb{delwx}
\end{figure}

  Then the Friedmann equation reads,
  \be
  \frac{H^2}{H_0^2}=\Omega_{X0}(1+z)^{3(1+w_X)}
 +\frac{1-\Omega_{X0}}{\delta+3w_X} \left[ \delta
(1+z)^{3(1+w_X)}+3w_X (1+z)^{3-\delta} \right],
  \en
 where $\Omega_{\rm X0}=\kappa^2\rho_{\rm X0}/(3H_0^2)$, and we have used
 $\Omega_{\rm dm0}+\Omega_{\rm X0}=1$. Thus we need to constrain the
 three parameters $\delta, \Omega_{\rm X0}, w_{\rm X}$. The
 constraint result by SNLS data is shown in fig 9. $\delta=0$ and
 $w_X=-1$ are indicated by the horizontal and vertical dashed lines,
 which represent the non-interacting XCDM model and interacting
 $\Lambda$CDM model, respectively.

 \begin{figure}
\begin{center}
\includegraphics[scale=.35]{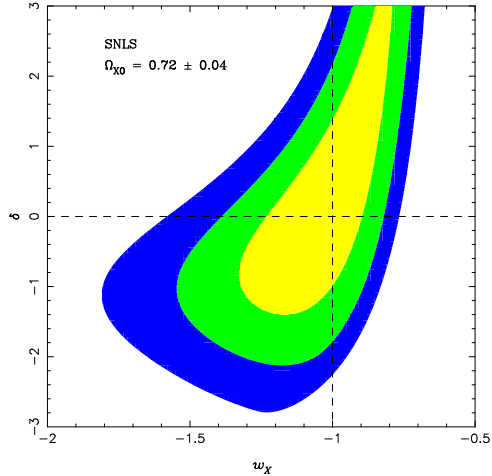}
\caption{Constraints of ($w_X,\delta$) by SNLS data at 68.3\%,
95.4\% and 99.7\% confidence levels marginalized over $\Omega_{X0}$
with priors $\Omega_{X0}=0.72 \pm 0.04$ and $\delta<3$. From
\cite{guointer}}
\end{center}
\lb{interXf}
\end{figure}

 From fig 8 and 9, we see that the
 observations leave enough space for the parameters ($\delta$, $w_X$)
 to cross the phantom divide.

  In the previous interacting model, we consider a phenomenological
  interaction, which is put in ``by hand". We should find a more
  sound
  physical foundation for the interactions. We will deduce an
  interaction term from the low energy limit of string/M theory in
  the scenario of the interacting ~\cp ~model \cite{negden1}.

  The \cp ~model was suggested as a
candidate of a unified model of dark energy
 and dark matter \cite{cp}.  The \cp~ is characterized by an exotic equation
of state
 \be
 p_{ch}=-A/\rho_{ch},
 \label{state}
 \en
 where $A$ is a positive constant. The above equation of state leads to
 a density evolution in the form
 \be
 \rho_{ch}=\sqrt{A+\frac{B}{a^6}},
 \label{evo}
 \en
 where $B$ is an integration constant.
The attractive feature of the model is that it naturally unifies
both dark energy and dark matter. The reason is that, from
(\ref{evo}), the \cp~ behaves as dust-like matter at early stage and
as a cosmological constant at later stage.

 Though \cp~has such a nice property, it is a serious flaw when one
 studies the fluctuation growth in \cp~model. It is found that \cp~ produces oscillations or exponential blowup of the matter
   power spectrum, which is inconsistent with
    observations \cite{antiudm}. So we turn to a model that the \cp~only plays the
   role of dark energy. To cross the phantom divide we consider a
   model in which the \cp~couples to dark mater.

       Although non-minimal coupling between the dark energy and ordinary
 matter fluids  is strongly restricted by the  experimental tests in
 the solar system \cite{will}, due to the unknown nature of the dark
 matter as part of the background, it is possible to have
 non-gravitational interactions between the dark energy and the dark
 matter components, without conflict with the experimental data. Thus, the observation constrain
 the only proper candidate to be coupled to \cp~is  dark matter.

 We consider the original \cp, whose pressure and energy density satisfy the
 relation, $p_{ch}=-A/\rho_{ch}$. By assuming the cosmological principle
 the continuity equations are written as
 \be
 \dot{\rho}_{ch}+3H \gamma_{ch} \rho_{ch}=-\Gamma,
 \label{1st conti}
 \en
 and
 \be
 \dot{\rho}_{dm}+3H \gamma_{dm} \rho_{dm}=\Gamma,
 \label{2nd conti}
 \en
 where the subscript $dm$ denotes dark matter,  and $\gamma$ is defined as
 \be
 \gamma=1+\frac{p}{\rho}=1+w,
 \en
 in which $w$ is the parameter of the state of equation,
  and $\gamma_{dm}=1$ throughout the evolution of the universe,
 whereas $\gamma_{ch}$ is a variable.

 $\Gamma$ is the
 interaction term between \cp~and dark matter. Since there does not
exist any microphysical hint on the possible nature of a coupling
between dark matter and \cp~(as dark energy), the interaction terms
between dark energy and dark matter are rather arbitrary in
literatures \cite{inter}. Here
 we try to present a possible origin from fundamental field theory for
 $\Gamma$.

 Whereas we are still lack of a
complete formulation of unified theory of all interactions
(including gravity, electroweak and strong), there at present is at
least one hopeful candidate, string/M theory. However, the theory is
far away from mature such that it is still not known in a way that
would enable us to ask the questions about space-time in a general
manner, say nothing of the properties of realistic particles.
Instead, we have to either resort to the effective action approach
which takes into account stringy phenomena in perturbation theory,
or we could study some special classes of string solutions which can
be formulated in the non-perturbative regime. But the latter
approach is available only for some special solutions, most notably
the BPS states or nearly BPS states in the string spectrum: They
seems to have no relation to our realistic Universe. Especially,
there still does not exist a non-perturbative formulation of generic
cosmological solutions in string theory. Hence nearly all the
investigations of realistic string cosmologies have been carried out
essentially in the effective action range. Note that the departure
of string-theoretic solutions away from general relativity is
induced by the presence of additional degrees of freedom which
emerge in the massless string spectrum. These fields, including the
scalar dilaton field, the torsion tensor field, and others, couple
to each other and to gravity non-minimally, and can influence the
dynamics significantly. Thus such an effective low energy string
theory deserve research to solve the dark energy problem. There a
special class of scalar-tensor theories of gravity is considered to
avoid singularities in cosmologies in \cite{st}. The action is
written below,
 \bea
 S_{st}=\int d^4x \sqrt{-g} \left[\frac{1}{16\pi G}R-
 \frac{1}{2}\partial_{\mu} \phi \partial^{\mu} \phi +
 \frac{1}{q(\phi)^2}
 L_{\rm dm}(\xi, \partial \xi, q^{-1}g_{\mu\nu})\right],
 \label{staction}
 \ena
  where $G$ is the Newton gravitational constant, $\phi$ is a scalar
  field, $L_{\rm dm}$ denotes Lagrangian of matter , $\xi$ represents
  different matter degrees of matter fields, $q$ guarantees the coupling strength
  between the matter fields and the dilaton. With action
  (\ref{staction}), the interaction term can be written as follow \cite{st},
  \be
  \Gamma=H\rho_{\rm dm} \frac{d\ln q'}{d\ln a}.
  \en
  Here we introduce new variable $q(a)'\triangleq
  q(a)^{(3w_n-1)/2}$, where $a$ is the scale factor in standard
  FRW metric.   By assuming
  \be
  q'(a)=q_0e^{3\int c(\rho_{\rm dm}+\rho_{\xi})/\rho_{\rm dm} d\ln a  },
  \en
  where $\rho_{\rm dm}$ and $\rho_{\xi}$ are the densities of
  dark matter and the scalar field respectively, one arrive at the interaction term,
  \be
  \Gamma=3Hc(\rho_{\rm dm}+\rho_\phi).
  \en

  With this interaction form we study the equation set (\ref{1st conti}) and
  (\ref{2nd conti}). Set
  $s=-\ln(1+z)$, $\Gamma=3Hc(\rho_{ch}+\rho_{dm})$,
  $u=(3H_0^2)^{-1}(3\mu^2)^{-1} \rho_{dm}$, $v=(3H_0^2)^{-1}(3\mu^2)^{-1}
  \rho_{ch}$, $A'=A(3H_0^2)^{-2}(3\mu^2)^{-2}$, where $c$ is a constant without dimension.
   Using these variables, (\ref{1st conti}) and
  (\ref{2nd conti}) reduce to
  \be
  \frac{du}{ds}=-3u+3c(u+v),
  \label{auto1}
  \en
  \be
  \frac{dv}{ds}=-3(v-A'/v)-3c(u+v).
  \label{auto2}
  \en
  We note that the variable time does not appear in the dynamical system
  (\ref{auto1}) and (\ref{auto2}) because time has been completely
  replaced by redshift $s=-ln(1+z)$. The critical points of
  dynamical system (\ref{auto1}) and (\ref{auto2}) are given by
  \be
  \frac{du}{ds}=\frac{dv}{ds}=0.
  \en
  The solution of the above equation is
  \bea
  u_c=\frac{c}{1-c}v_c,
 \label{uc}
  \\
  v_c^2=(1-c)A'.
  \label{vc}
  \ena
  We see the final state of the model contains both \cp ~and dark matter
  of constant densities if the singularity is stationary. The final
  state satisfies perfect cosmological principle: the universe is
   homogeneous and isotropic in space, as well as constant in time.
   Physically $\Gamma$ in (\ref{2nd conti}) plays the role of matter
   creation term $C$ in the theory of steady state universe at the
   future time-like infinity.
  Recall that $c$ is the coupling constant, may be positive or negative,
  corresponds the energy to transfer from \cp~ to dark matter or
  reversely. $A'$ must be a positive constant, which denotes the
  final energy density if $c$ is fixed. Also we can derive an
  interesting and simple relation between the static energy density
  ratio
  \be
  c=\frac{r_s}{1+r_s},
  \en
  where
  \be
  r_s=\lim_{z\to -1}\frac{\rho_{dm}}{\rho_{ch}}.
  \en
  To investigate the properties of the dynamical system in the
  neighbourhood of the singularities, impose a perturbation
  to the critical points,
  \bea
  \frac{d(\delta u)}{ds}=-3\delta u+3c(\delta u+\delta v),
  \label{linear1}
  \\
  \frac{d(\delta v)}{ds}=-3(\delta v+\frac{A'}{v_c^2}\delta v)-3c(\delta
  u+\delta v).
  \label{linear2}
  \ena
  The eigen equation of the above linear dynamical system $(\delta u,
  \delta v)$ reads
  \be
  (\lambda/3)^2+(2+\frac{1}{1-c})\lambda/3+2-2c^2=0,
  \label{eigen}
  \en
  whose discriminant is
  \be
  \Delta=[(1-c)^4+(3/2-c)^2]/(1-c)^2\geq 0.
  \en
  Therefore both of the two roots of eigen equation (\ref{eigen}) are real, consequently
    centre and focus singularities can not appear.
  Furthermore only $r_s\in (0,\infty)$, such that $c\in (0,1)$, makes
  physical sense. Under this condition it is easy to show that both
  the two roots of (\ref{eigen}) are negative. Hence the two
  singularities are stationary. However it is only the property
  of the linearized system (\ref{linear1}) and (\ref{linear2}),
  or  the property of orbits of the neighbourhoods
  of the singularities, while global Poincare-Hopf theorem requires that the
  total index of the singularities equals the Euler number of the
  phase space for the non-linear system (\ref{auto1}) and (\ref{auto2}). So there exists other
  singularity except for the two nodes. In fact it is a non-stationary saddle point at
  $u=0,~v=0$ with index $-1$. This singularity has been omitted in solving
  equations (\ref{auto1}) and (\ref{auto2}). The total index of the three singularities
  is $1$, which equals the Euler number of the phase space of this plane dynamical
  system. Hence there is no other singularities in this system.
    From these discussions we conclude that the global outline of
  the orbits of this non-linear dynamical system (\ref{auto1}) and (\ref{auto2})
   is similar to the electric fluxlines of two
  negative point charges.
    Here we plot figs 10 and 11 to show the properties of evolution of
    the universe controlled by the dynamical system (\ref{auto1}) and
    (\ref{auto2}). As an example we set $c=0.2,~~A'=0.9$ in figs 10-12.

 \begin{figure*}
\centering
\includegraphics[totalheight=2.2in, angle=0]{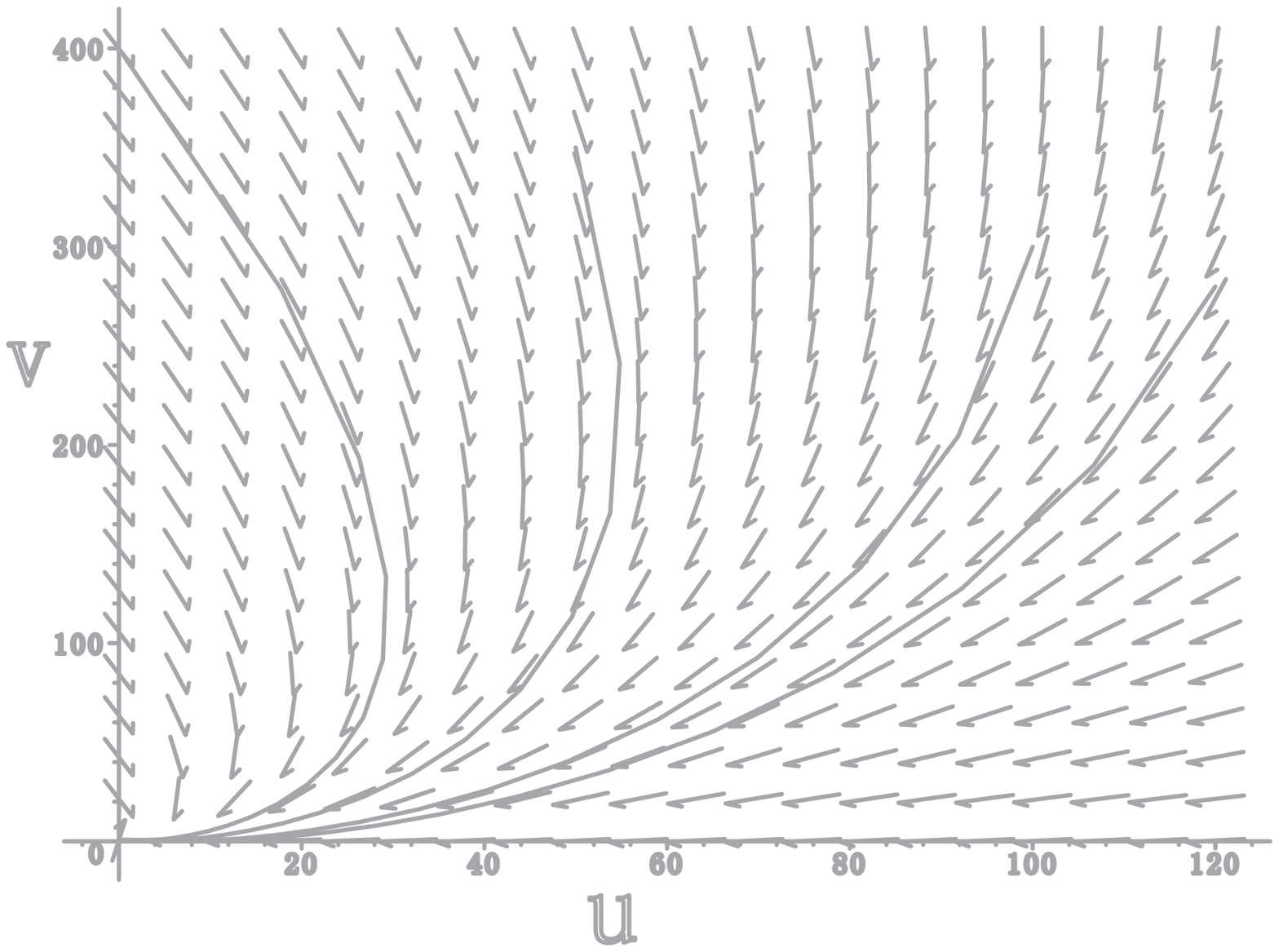}
\includegraphics[totalheight=2.2in, angle=0]{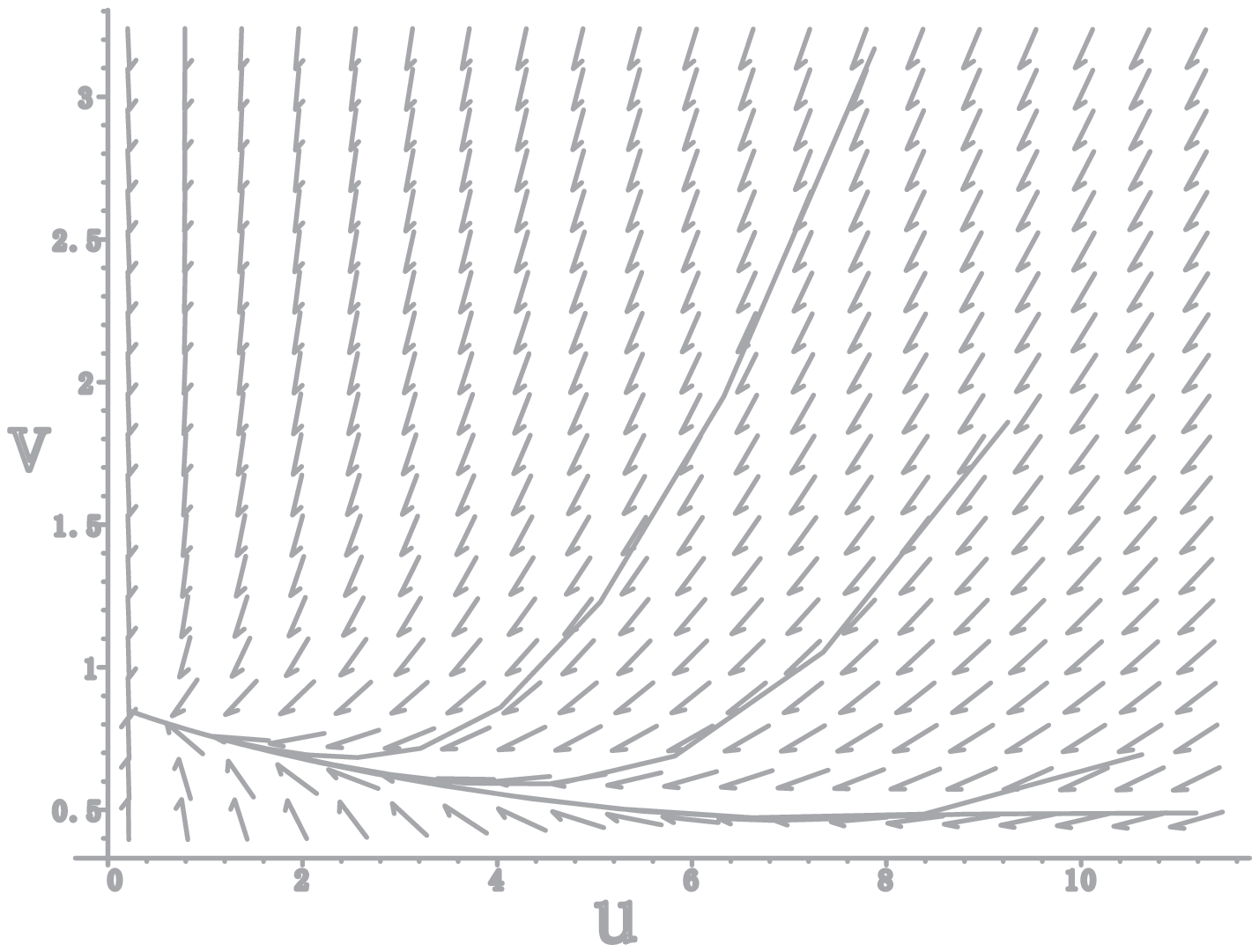}
\caption{The plane v versus u. {\bf{(a)}} left panel: We consider
the evolution of the universe from redshift $z=e^2-1$. The initial
condition is taken as $u=0,~v=400$; $u=50,~v=350$; $u=100,~v=300$;
$u=120,~v=280$ on the four orbits, from the left to the right,
respectively. It is clear that there is a stationary node, which
attracts most orbits in the first quadrant. At the same time the
orbits around the neibourhood of the singularity is not shown
clearly. {\bf{(b)}} right panel: Orbit distributions around the node
$u_c=v_c c/(1-c),~v_c=\sqrt{(1-c)A'}$. From \cite{negden1}}
 \label{globalgreen}
 \end{figure*}

\begin{figure}
\centering
\includegraphics[totalheight=3in, angle=-90]{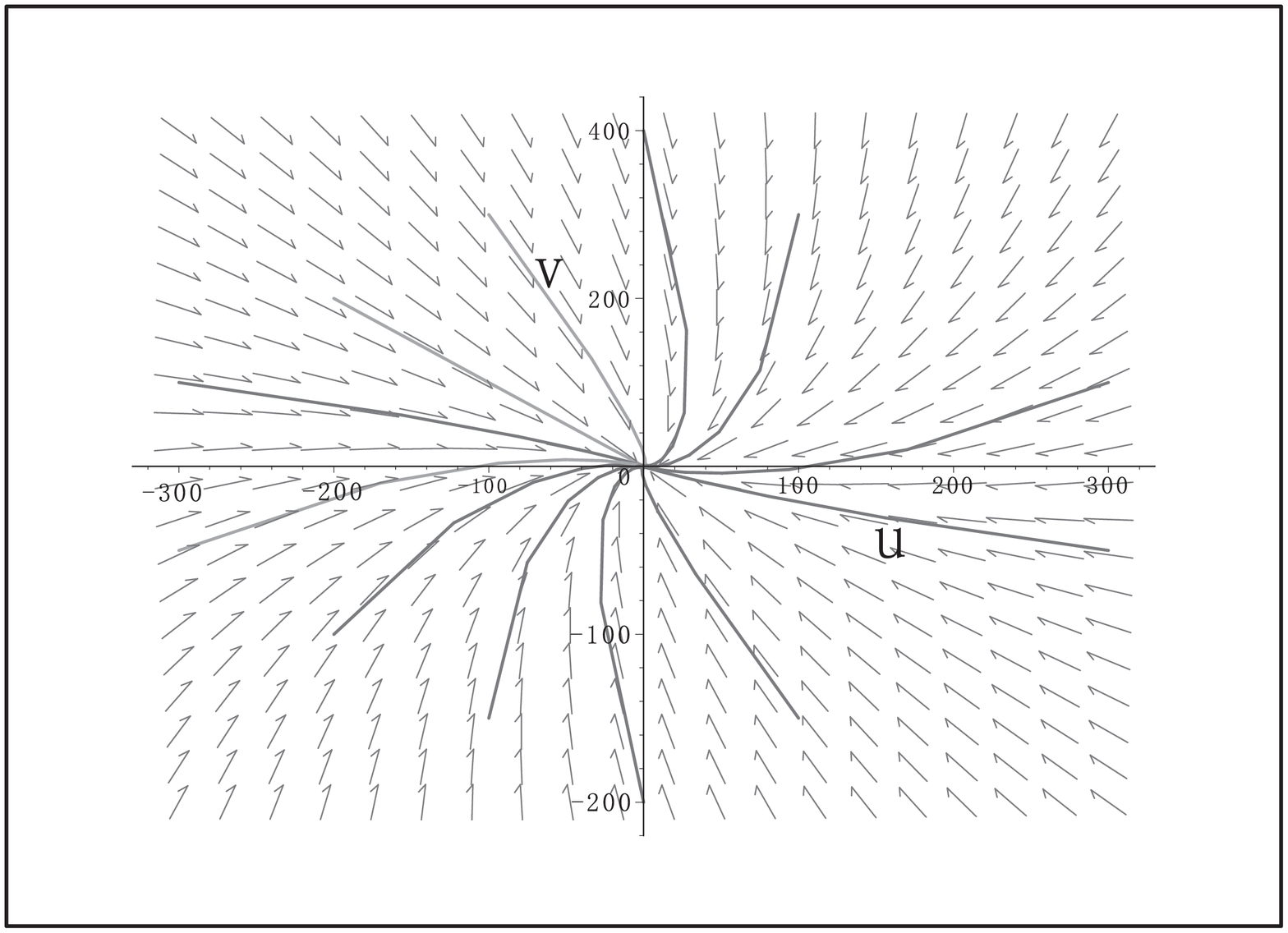}
\includegraphics[totalheight=3in, angle=-90]{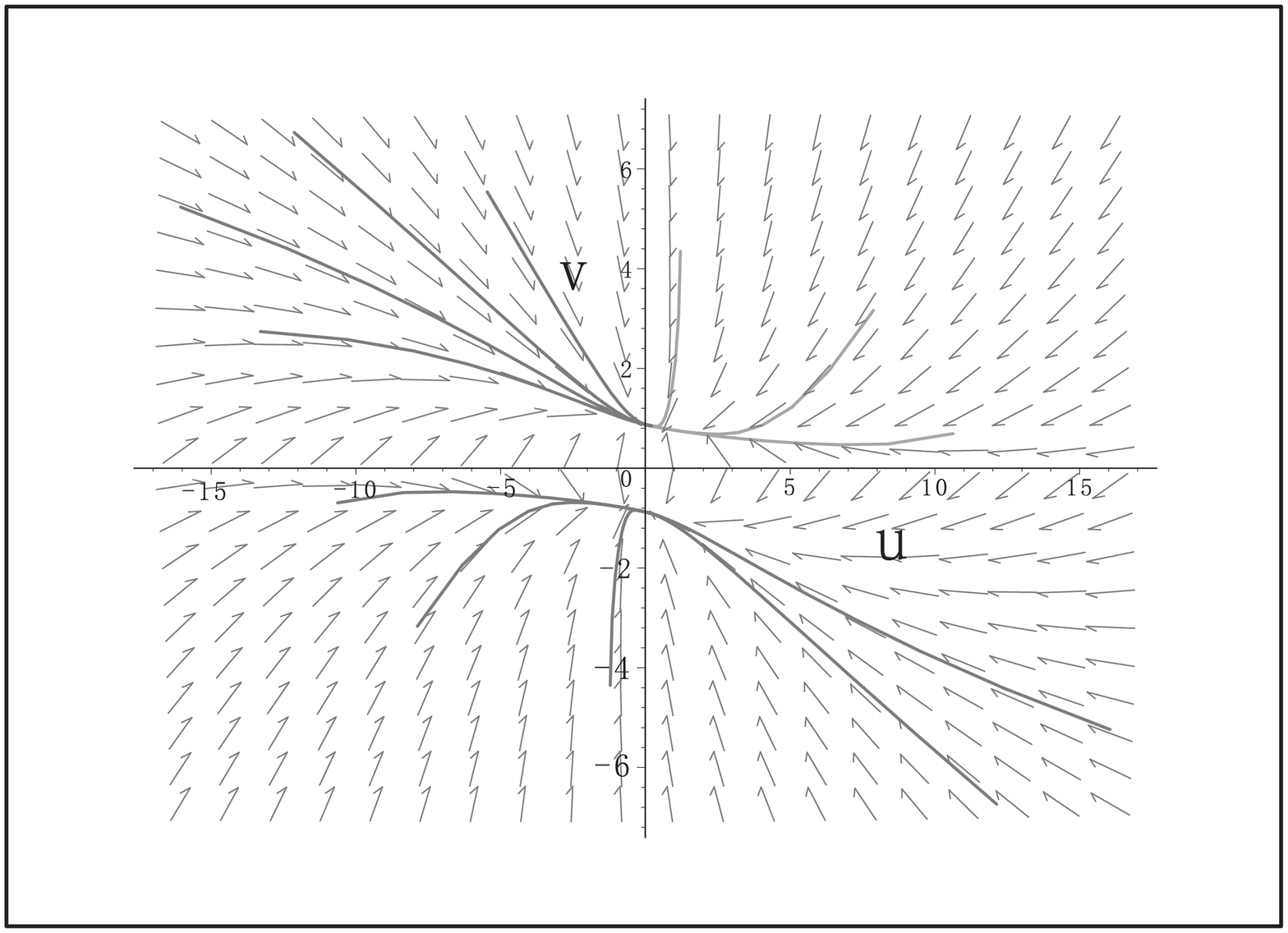}
\caption{The plane v versus u. {\bf{(a)}} left panel:    To show the
global properties of dynamical system (\ref{auto1}) and
(\ref{auto2}) we have to include some ``unphysical '' initial
conditions, such as $u=-100,~v=-300$, except for physical initial
conditions which have been shown in figure \ref{globalgreen}.
{\bf{(b)}} right panel: Orbits distributions around the nodes. The
two nodes
 $u_c=v_c c/(1-c),~v_c=\sqrt{(1-c)A'}$ and $u_c=v_c c/(1-c),~v_c=-\sqrt{(1-c)A'}$
 keep reflection symmetry about the original point. Just as we have analyzed, we see
 that the orbits of this dynamical system are similar to the electric fluxlines of two
  negative point charges. From \cite{negden1}}
 \label{global}
 \end{figure}

\begin{figure}
 \centering
 \includegraphics[totalheight=2.5in, angle=0]{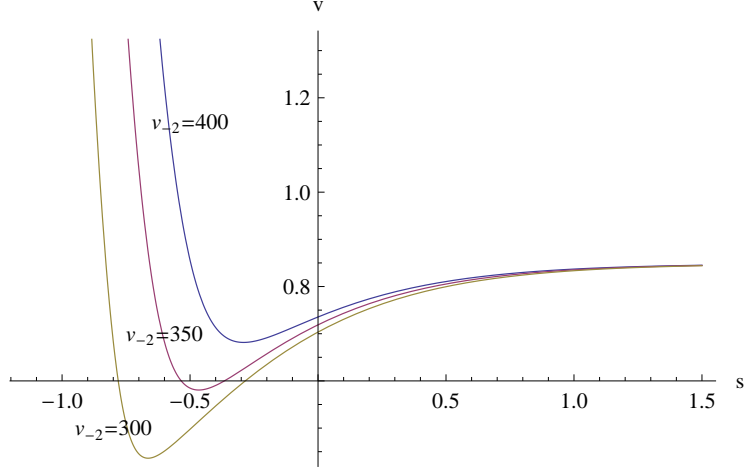}
 \caption{v versus s. The evolution of $v$ with different initial conditions
 $u(-2)=0,~v(-2)=400$;
  $u(-2)=50,~v(-2)=350$; $u(-2)=100,~v(-2)=300$ reside on the blue, red, and yellow curves, respectively. Obviously the energy density
  of \cp~  rolls down and then climbs up in some low redshift region. So the \cp
  ~dark energy can cross the phantom divide $w=-1$ in a fitting where the
  dark energy is treated as an independent component to dark matter. From \cite{negden1}, this figure has been re-plotted.   }
 \label{figv}
 \end{figure}

  Further, to compare with observation data we need the explicit
  forms of $u(x)$ and $v(x)$, especially $v(x)$. We need
  the properties of $\gamma_{ch}$ in our model, which is contained in
  $v(x)$, to compare with observations. Eliminate $u(x)$ by using
  (\ref{auto1}) and (\ref{auto2}) we derive
  \bea
  \frac{1}{3c}\frac{d^2v}{ds^2}+[1+(1+A'/v^2)/c]\frac{dv}{ds}+3cv+
  3(1-c)\left\{v+\left[\frac{dv}{ds}+3(v-A'/v)\right]/(3c)\right\}=0,
  \ena
  which has no analytic solution. We show some numerical solutions
  in figure \ref{figv}. We find that for proper region of parameter
  spaces, the effective equation of state of \cp~ crosses the
  phantom divide successfully.

  Up to now all of our results do not depend on Einstein field
  equation. They only depend on the most sound principle in physics, that is, the
  continuity principle, or the energy conservation law.  Different gravity theories
    correspond to different constraints imposed on our
   previous discussions.
    Our improvements show how far we can reach without information of dynamical
    evolution of the universe.

\begin{figure}
 \centering
 \includegraphics[totalheight=2.6in, angle=0]{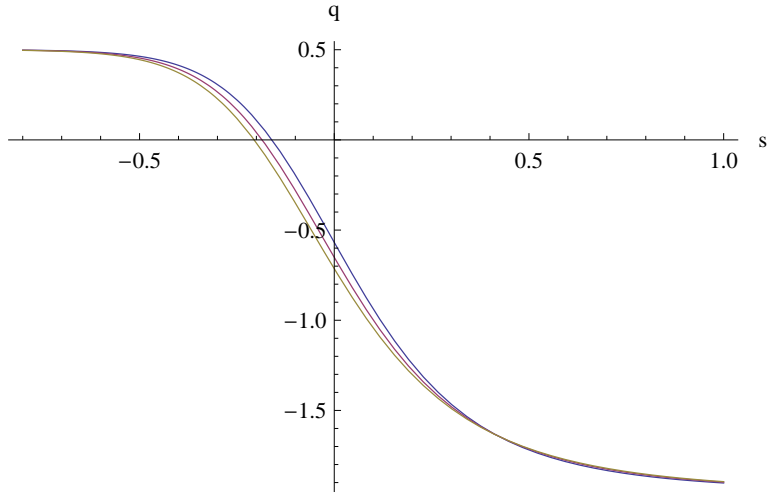}
 \caption{q versus s.
 The evolution of $q$ with different initial conditions
 $u(-2)=0,~v(-2)=273$;
  $u(-2)=15,~v(-2)=250$; $u(-2)=25,~v(-2)=233$, reside on the blue, red, and yellow curves, respectively. Evidently the
  deceleration parameter $q$
  of \cp~  rolls down and crosses $q=0$ in some low redshift region. The transition
  from deceleration phase to acceleration phase occurs at $z=0.18;~z=0.21;~z=0.23$
  to the curves $u(-2)=0,~v(-2)=273$;
  $u(-2)=15,~v(-2)=250$; $u(-2)=25,~v(-2)=233$, respectively.  One
  finds $-q\thickapprox 0.5\sim 0.6$ at $z=0$, which is well consistent with  observations. From \cite{negden1}, this figure has been re-plotted.  }
 \label{dece}
 \end{figure}

 (\ref{auto1}) illuminates that the dark matter in this interacting model does not behaves as dust. Qualitatively,
 the dark matter gets energy from dark energy for a positive $c$, and becomes soft, ie, its energy density
 decreases slower than $(1+z)^{3}$ in an expanding universe. The parameter which carries the total
  effects of cosmic fluids is the deceleration parameter $q$. From now on we introduce the Friedmann
  equation of the standard general relativity.
  As a simple case we study the evolution of $q$ in
  a spatially flat universe. So $q$ reads

  \be
  q=-\frac{\ddot{a}a}{\dot{a}^2}=\frac{1}{2}\left(\frac{u+v-3A'/v^2}
  {u+v}\right),
  \en
  and density of \cp~ $u$ and density of dark matter $v$ should
  satisfy
  \be
  u(0)+v(0)=1.
  \label{con}
  \en
  And then Friedmann equation ensures the spatial flatness in the
  whole history of the universe.
  Before analyzing the evolution of $q$ with redshift,
   we first study its asymptotic behaviors. When $z\to \infty$, $q$
 must go to $1/2$ because both \cp~ and dark matter behave like dust
 , while when $z\to -1$ $q$ is determined by
 \be
  \lim_{z\to -1} q=\frac{1}{2}\left(\frac{u_c+v_c-3A'/v_c^2}
  {u_c+v_c}\right).
  \en
 One can finds the parameters $c=0.2,~A'=0.9$ are difficult to
  content the previous constraint Friedmann constraint ($\ref{con}$). Here we carefully
  choose a new set of parameter which satisfies Friedmann constraint
  (\ref{con}), say, $A'=0.4,~c=0.06$. Therefore we obtain
 \be
 \lim_{z\to -1} q=-1.95,
 \en
  by using (\ref{uc}) and (\ref{vc}).
  Then we plot figure \ref{dece} to clearly display the evolution of
  $q$. One can check $u(0)=0.25,~v(0)=0.75;~u(0)=0.28,~v(0)=0.72;
  ~u(0)=0.3,~v(0)=0.7$, respectively on the curves
  $v(-2)=273;~v(-2)=250;~v(-2)=233$. One may find an interesting
  property of the deceleration parameter displayed in fig
  \ref{dece}: the bigger the proportion of the dark energy, the
  smaller the absolute value of the deceleration parameter. The
  reason roots in the extraordinary state of \cp~(\ref{state}), in
  which the pressure $p_{ch}$ is inversely proportional to the energy density $\rho_{ch}$.

 Also we note that maybe an FRW universe with non-zero
  spatial curvature fits  deceleration parameter
  better than spatially flat FRW universe. This point deservers to
  research further.

  After the presentation of the  original interacting \cp~ model, there are several
  generalizations. For details of these generalizations, see
  \cite{generalcp}.

 \subsection{model in frame of modified gravity}

  The judgement that there exists an exotic component with negative
  pressure, or dark energy, which accelerates the universe, is
  derived in frame of general relativity. The validity of general
  relativity has been well tested from the scale of millimeter to
  the scale of the solar system. Beyond this scale, the evidences
  are not so sound. So we should not be surprised if general relativity
  fails at the scale of the Hubble radius. Surely, any new gravity
  theory must reduce to general relativity at the scale between
  millimeter to the solar system. In frame of the new gravity
  theories, the cosmic acceleration may be a natural result even we only
  have dust in the universe.

  There are various suggestions on how to modify general relativity.
  In this brief review we concentrates on the brane world theory. Inspired by the developments of string/M theory, the idea that our
universe is a 3-brane embedded in a higher dimensional spacetime has
received a great deal of attention in recent years. In this brane
world scenario, the standard model particles are confined on the
3-brane, while the gravitation can propagate in the whole space. In
this picture, the gravity field equation gets modified at the left
hand side (LHS) in (\ref{ein}), while the dark energy is a stuff put
at the right hand side (RHS) in (\ref{ein}). In the modified gravity
model, the surplus geometric terms respective to the Einstein tensor
play the role of the dark energy in general relativity.

We consider a 3-brane imbedded in a 5-dimensional bulk.
 The action includes the action of the bulk and the action of the brane,
 \be
 \label{action}
 S=S_{\rm bulk}+S_{\rm brane}.
 \en
 Here
 \be
 \label{actionbulk}
  S_{\rm bulk} =\int_{\cal M} d^5X \sqrt{-{g_5}}
{\cal L}_{\rm bulk},
 \en
  where $X=(t,z,x^1,x^2,x^3)$ is the bulk coordinate, $x^1,x^2,x^3$ are the coordinates of the maximally symmetric space. ${\cal
  M}$  denotes the bulk manifold. The bulk Lagrangian can be
 \be
   {\cal L}_{\rm bulk}={1 \over 2\kappa_5^2 }\left[ R_5 +\alpha F(R_5)\right]+{\cal L}_{\rm
   m}+\Lambda_5,
   \lb{5dimlag}
  \en
 where  ${g_5}$, $\kappa_5$, $R_5$, ${\cal L}_{\rm m}$,
  denote the bulk manifold, the determinant of the
  bulk metric,  the 5-dimensional Newton constant,
   the 5-dimensional Ricci scalar, and the
  bulk matter Lagrangian, respectively. $F(R_5)$ denotes the higher
  order term of scalar curvature $R_5$, the Ricci curvature $R_{5{\rm AB}}$, the Riemann curvature $R_{5{\rm ABCD}}$.

  There are too much possibilities and rather arbitrary to choose
  the higher order terms. Generally
 the resulting equations of motion of such a term give more than second
derivatives of metric and the resulting theory is plagued by ghosts.
However there exists a combination of quadratic terms, called
Gauss-Bonnet term, which generates equation of motion without the
terms more than second derivatives of metric and the theory is free
of ghosts \cite{earlygb}.  Another important property of
Gauss-Bonnet term is that, just like Hilbert
 Lagrangian is a pure divergence in
 2 dimensions and Einstein tensor identifies zero in 1 and
 2 dimensions, we have that in 4 or less dimension the Gauss-Bonnet
 Lagrangian is a pure divergence. We see the dilemma of quadratic
 term in 4 dimensional theory: if we include it with  non pure
 divergence we shall confront ghosts; if we want to remove
 ghosts we get a pure divergence term. So only in theories in more
 than 4 dimensional Gauss-Bonnet combination provides physical
 effects. Moreover the Gauss-Bonnet term also appears in both
 low energy effective action of Bosonic string theory
 \cite{stringgb1} and low energy effective action of
 Bosonic modes of heterotic and type II super string theory
  \cite{stringgb2}. An investigation into the effects of
  a Gauss-Bonnet term in the 5 dimensional bulk of brane world
 models is therefore well motivated. The Gauss-Bonnet term in 5
 dimension reads,

  \be
  F(R_5)= R_5^{2}
  -4R_{5{\rm AB}}
  R_5^{\rm AB}+R_{5{\rm ABCD}}R_5^{\rm ABCD}.
  \en

 The action of the brane can be written as,
 \be
 \label{actionbrane}
 S_{\rm brane}=\int_{M} d^4 x\sqrt{-g} \left(
{\kappa_5^{-2}} K + L_{\rm brane} \right),
 \en
 where  $M$
  indicates the brane manifold, $g$ denotes the determinant of the
  brane metric, $L_{\rm brane}$ stands for the Lagrangian confined to the brane,
    and $K$ marks the trace of the second fundamental form of the
    brane. $x=(\tau, x^1,x^2,x^3)$ is the brane coordinate. Note
    that $\tau$ is not identified with $t$ if the the brane is not
    fixed at a position in the extra dimension $z=$constant. We will
    investigate the cosmology of a moving brane along the extra dimension
    $z$ in the bulk, and such that $\tau$ is different from $t$.

    We set the Lagrangian confined to the brane as follows,
     \be
    \label{4dimlag}
    L_{\rm brane}=\frac{1}{16\pi G}R-\lambda + L_{\rm
m},
  \en
  where $\lambda$ is the brane tension and $L_{\rm m}$ denotes the
  ordinary matter, such as dust and radiation, located at the brane.
  $R$ denotes the 4 dimensional scalar
curvature term on the brane, which is an important one except a
Gauss-Bonnet term in
 the bulk.  This induced gravity correction arises because the
localized matter fields on the brane, which couple to bulk
gravitons, can generate via quantum loops a localized
four-dimensional world-volume kinetic term for gravitons~\cite{dgp}.

 Assuming there is a  mirror symmetry in the bulk, we
  have the Friedmann equation on the brane \cite{combinecos},
  see also \cite{cov},
  \bea
 {4\over r_c ^2}\left[1 +\frac{8}{3}\alpha\left(H^2 +{k \over a^2}
 + {U \over 2} \right) \right]^{2}\left(H^2 +{k \over
 a^2}-U\right) =\left(H^2 +{k \over a^2} -\frac{8\pi G} {3}(\rho+\lambda)\right)^{2},
 \label{fried3}
 \ena
 where
 \be
 \label{U}
  U=-\frac{1}{4\alpha}\pm \frac{1}{4\alpha} \sqrt{1+
  4\alpha\left(\frac{\Lambda_5}{6}+\frac{M\kappa_5^2}{4 \pi
  a^4}\right)},
  \en
 \be r_c= \kappa_5^2\mu^2. \en
  Here $M$ is a constant, standing for the mass of bulk black
  hole. For various limits of (\ref{fried3}), see \cite{selfgbde}.

  For convenience, we introduce the
 following new variables and parameters,
 \begin{eqnarray}
 \label{parameters}
 && x \equiv \frac{H^2}{H_0^2}+\frac{k}{a^2H_0^2}=
   \frac{H^2}{H_0^2}-\Omega_{k0} (1+z)^2,  \nonumber \\
 && u \equiv \frac{8\pi \ ^{(4)}G}{3H_0^2}(\rho +\lambda)=
 \Omega_{m0}(1+z)^3+\Omega_{\lambda}, \nonumber \\
 && m \equiv \frac{8}{3} \alpha H_0^2, \nonumber
\\
&&  n \equiv \frac{1}{H_0^2r_c^2}, \nonumber \\
 &&
 y  \equiv \frac{1}{2}UH_0^{-2}=\frac{1}{3m}\left(-1+\sqrt{1+
 \frac{4\alpha \Lambda_5}{6}+{\frac{8\alpha M
 {}^{(5)}G}{a^4}}}\right) \nonumber \\
 &&~~~=
 \frac{1}{3m}\left(-1+\sqrt{1+m\Omega_{\Lambda_5}+m\Omega_{M0}(1+z)^4}\right),
 \end{eqnarray}
 and we have assumed that there is only pressureless dust in the universe.  As before,
 we have used the following notations
 \begin{equation}
 \label{Omega}
\Omega_{k0}=-\frac{k}{a_0^2H_0^2},~~\Omega_{m0}=\frac{8\pi
  G}{3}\frac{\rho_{m0}}{H_0^2},~~\Omega_{\lambda}=\frac{8\pi
  G}{3}\frac{\lambda}{H_0^2},
  ~~\Omega_{\Lambda_5}=
  \frac{3\Lambda_5}{8H_0^2},~~\Omega_{M0}=\frac{3M
  \kappa_5^2}{8\pi a_0^4H_0^2}.
  \end{equation}
With these new
 variables and parameters,
 (\ref{fried3}) can be rewritten as
 \be
 \label{GBI}
 4n(x-2y)[1+m(x+y)]^2=(x-u)^2.
 \en
 This is a cubic equation of the variable $x$. According to algebraic theory it
 has 3 roots. One can explicitly write down three roots. But they are too lengthy and
 complicated to present here. Instead we only express those three roots formally
 in the order given in {\it Mathematica}
 \begin{eqnarray}
 \label{solution}
 && x_1= x_1(y,u|m,n), \nonumber \\
 && x_2=x_2(y,u|m,n), \nonumber \\
&& x_3=x_3(y,u|m,n),
\end{eqnarray}
where $y$ and $u$ are two variables, $m$ and $n$ stand for two
parameters. The root on $x$ of the equation (\ref{GBI}) gives us the
modified Friedmann equation on the Gauss-Bonnet brane world with
induced gravity. From the solutions given in (\ref{solution}), this
model seems to have three branches.
 In addition,
note that all parameters introduced in (\ref{parameters}) and
(\ref{Omega}) are not independent of each other. According to the
Friedmann equation (\ref{solution}), when all variables are taken
current values, for example, $z=0$, the Friedmann equation will give
us a constraint on those parameters,
\begin{equation}
\label{constraint}
1=f(\Omega_{k0},~\Omega_{m0},~\Omega_{M0},~\Omega_{\Lambda_5},
  ~\Omega_{\lambda},~m,~n).
\end{equation}

 To compare with observation, we introduce the concept ``equivalent
 dark energy" or ``virtual dark energy" in the modified gravity
 models, since almost all the properties of dark energy are deduced
 in the frame of general relativity with a dark energy.

 The Friedmann equation in the
four dimensional
  general relativity can be written as
 \be
 H^2+\frac{k}{a^2}=\frac{8\pi G}{3} (\rho+\rho_{de}),
 \label{genericF}
 \en
 where the first term of RHS of the above equation represents the dust matter and the second
 term stands for the dark energy.
 Generally speaking the Bianchi identity requires,
  \be
 \frac{d\rho_{de}}{dt}+3H(\rho_{de}+p_{de})=0,
 \label{em}
 \en
 we can then express the equation of state for the dark
 energy as
   \be
  w_{de}=\frac{p_{de}}{\rho_{de}}=-1-\frac{1}{3}\frac{d \ln \rho_{de}}{dlna}.
  \label{wde}
   \en
  Note that we can rewrite the Friedmann equation (\ref{solution}) in the form of
 (\ref{genericF}) as
 \be
 \label{xh^2}
 xH_0^2=\frac{8\pi G}{3} \rho+\left(H_0^2 x(y,u|m,n)-\frac{8\pi G}{3}
 \rho\right)=\frac{8\pi G}{3}( \rho+Q),
 \en
 where $\rho$ is the energy density of dust matter on the brane
 and the term
 \be
 Q \equiv \frac{3H_0^2}{8\pi G} x(y,u|m,n)-
 \rho
 \en
 corresponds to $\rho_{de}$ in (\ref{genericF}).

\begin{figure}
 \centering
 \includegraphics[totalheight=2.5in]{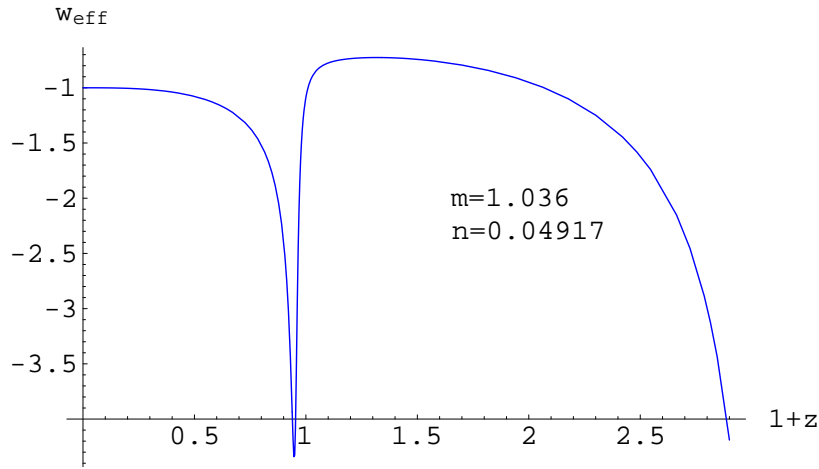}
 \caption{The equation of state $w_{eff}$ with respect to the
 red shift $1+z$, with $\Omega_{m0}=0.28$ and  $\Omega_A=2.08$. From \cite{selfgbde}.}
 \label{wm03}
 \end{figure}

 In fig \ref{wm03} we show the equation of state for the virtual dark
 energy when we take $m=1.036$ and $n=0.04917$. In this case, from
 the constraint equation (\ref{constraint}), one has
 $\Omega_{M0}=2.08$. From the figure we see that $w_{eff}<-1$ at
 $z=0$ and
 $$ \left. \frac{dw_{eff}}{dz}\right|_{z=0}<0.
 $$
Therefore the equation of state for the virtual dark energy can
indeed cross the phantom divide $w=-1$ near $z \sim 0$.

\begin{figure}
 \centering
 \includegraphics[totalheight=2.5in]{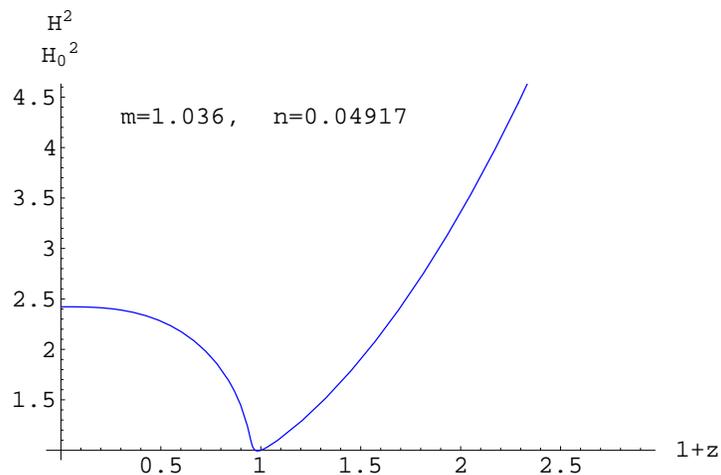}
 \caption{$H^2/H_0^2$ versus $1+z$,  with $\Omega_{m0}=0.28$
 and $\Omega_A=2.08$. From \cite{selfgbde}.}
 \label{hm03}
 \end{figure}

 Fig \ref{wm03} illuminates that the behavior of the virtual dark energy seems rather strange.
 However we should remember that it is only virtual dark energy, not
 actual stuff. The whole evolution of the universe is described by
 the Hubble parameter. We plot the Hubble parameter $H$
 corresponding to fig \ref{wm03} in fig \ref{hm03}.

 Fig \ref{hm03} displays that the universe will eventually becomes a de Sitter
 one. For more figures with different parameters, see
 \cite{selfgbde}. The constraint of this brane model with induced scalar term on the brane and Gauss-Bonnet term
 in the bulk has been investigated in \cite{DGPGBcon}.

 The above is an example of ``pure geometric" dark energy. We can
 also consider some mixed dark energy model, ie, the cosmic acceleration is
 driven by an exotic matter and some geometric effect in part. Why such an apparently
 complicated suggestion? There are many interesting models are
 proposed to explain the cosmic acceleration, including dark energy
 and modified gravity models. However, several influential and hopeful models, such as quintessence
 and DGP model, fundamentally can not account for the crossing $-1$
 behavior of dark energy. By contrast, some hybrid model of the dark energy and
 modified gravity may realize such a crossing. As an example we
 study the quintessence and phantom in frame of DGP \cite{self3}.

  Our starting point is still action (\ref{action}). In a DGP model
  with a scalar, (\ref{5dimlag}) becomes a pure Einstein-Hilbert action,

 \be
   {\cal L}_{\rm bulk}={1 \over 2\kappa_5^2 } R_5,
   \lb{5dimlagsim}
  \en
 and we add a scalar term in (\ref{4dimlag}),
  \be
    \label{4dimlagsim}
    L_{\rm brane}=\frac{1}{16\pi G}R + L_{\rm
m}+L_{\rm scalar}.
  \en
 Here the scalar term can be ordinary scalar (quintessence) or phantom (scalar with
 negative kinetic term).
 The Lagrangian of a quintessence reads,
 \be
 L_{\phi}=-\frac{1}{2}\partial_\mu \phi
  \partial^\mu \phi-V(\phi),
  \en
 and for phantom,
 \be
  L_{\psi}=\frac{1}{2}\partial_\mu \psi
  \partial^\mu \psi-U(\psi).
  \lb{lagphan}
  \en
   In an FRW universe we have
     \bea
  \rho_{\phi}=\frac{1}{2}\dot{\phi}^2+V(\phi),\\
  p_{\phi}=\frac{1}{2}\dot{\phi}^2-V(\phi).
  \ena

  The exponential potential is an important
 example which can be solved exactly in the standard model. Also it
 has been shown that the inflation driven by a scalar with
  exponential potential can exit naturally in the warped DGP model \cite{hs1}.
      It is therefore quite interesting to
 investigate a scalar with such a potential in late time universe on a DGP brane.
 Here we set
 \be
 V=V_0e^{-\lambda_1\frac{\phi}{\mu}}.
 \en
 Here $\lambda_1$ is a constant and $V_0$ denotes the initial value of
 the potential.

 The Friedmann equation
  (\ref{fried3})  becomes
  \bea
 H^2+\frac{k}{a^2}=\frac{1}{3\mu^2}\left[\rho+\rho_0+\theta\rho_0
 (1+\frac{2\rho}{\rho_0})^{1/2}\right],
 \label{fried4}
 \ena
 where
 \be
 \rho_0=\frac{6\mu^2}{r_c^2}.
 \en

  Similar to the previous case,  we derive
  the virtual dark energy by comparing (\ref{fried4}) and
  (\ref{genericF}),
  \be
 \rho_{de}=\rho_{\phi}+\rho_0+\theta\rho_0\left[\rho+\rho_0+\theta\rho_0
 (1+\frac{2\rho}{\rho_0})^{1/2}\right].
 \label{rhode2}
 \en

 From (\ref{wde}), we calculate the derivation of effective density of dark energy with respective to $\ln (1+z)$
 for a ordinary scalar,
   \bea
  \frac{d \rho_{de}}{d \ln
  (1+z)}=3[\dot{\phi}^2
  +\theta(1+\frac{\dot{\phi}^2+2V+2\rho_{dm}}{\rho_0})
  ^{-1/2}(\dot{\phi}^2+\rho_{dm})].
  \label{derivation or}
  \ena
  If $\theta=1$, both terms of RHS are positive, hence it never goes
  to zero at finite time. But if $\theta=-1$, the two terms of RHS
  carry opposite sign, therefore it is possible that the EOS of dark
  energy crosses phantom divide. In a scalar-driven DGP,
  we only consider the case of $\theta=-1$.

  For convenience,  we  define some
  dimensionless variables,
  \bea
  y_1&\triangleq&\frac{\dot{\phi}}{\sqrt{6}\mu H},\\
  y_2&\triangleq&\frac{\sqrt{V}}{\sqrt{3}\mu H},\\
  y_3&\triangleq&\frac{\sqrt{\rho_m}}{\sqrt{3}\mu H},\\
  y_4&\triangleq&\frac{\sqrt{\rho_0}}{\sqrt{3}\mu H}.
  \ena
  The Friedmann equation (\ref{fried4}) becomes
  \be
 \label{constraint}
 y_1^2+y_2^2+y_3^2+y_4^2-y_4^2\left(1+2\frac{y_1^2+y_2^2+y_3^2}{y_4^2}\right)^{1/2}=1.
 \en
 The stagnation point, that is, $d\rho_{de}/d\ln(1+z)=0$ dwells at
  \be
 \frac{y_4}{\sqrt{2}+y_4}\left(2+\frac{y_3^2}{y_1^2}\right)=2,
 \en
 which can be derived from (\ref{derivation or}) and
 (\ref{constraint}).

\begin{figure}
\centering
 \includegraphics[totalheight=1.8in, angle=0]{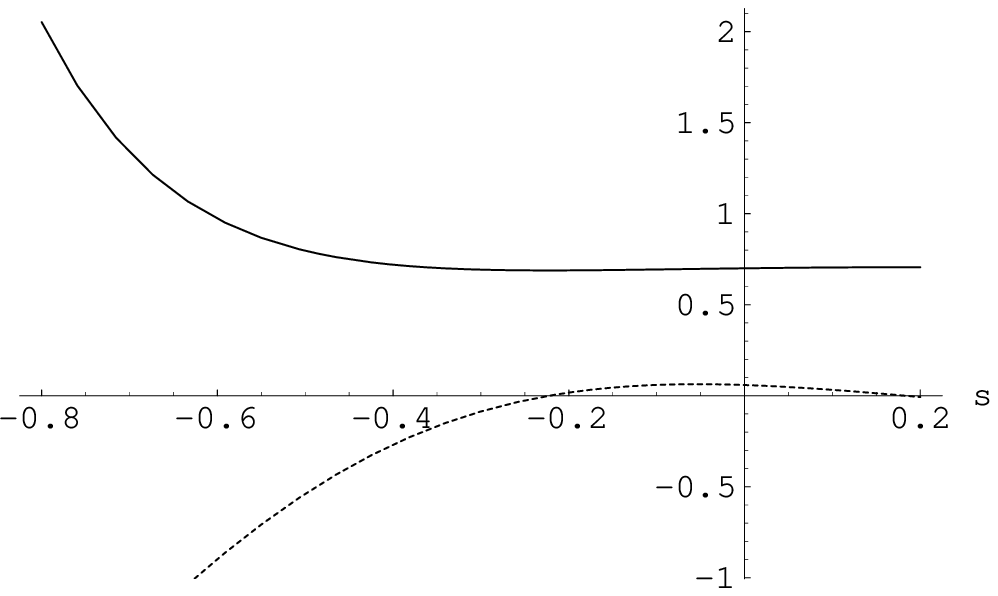}
\includegraphics[totalheight=2in, angle=0]{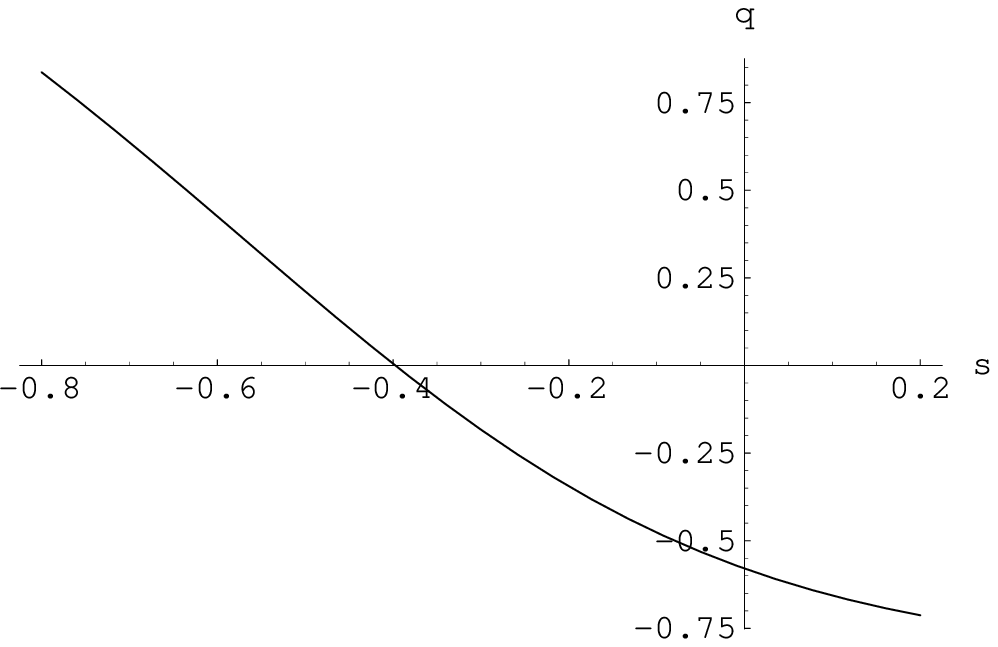}
\caption{For this figure, $\Omega_{ki}=0.01$, $\Omega_{r_c}=0.01$,
$\lambda_1=0.5$. {\bf{(a)}} The left panel: $\beta$ and $\gamma$ as
functions of $s$, in which $\beta$ resides on the solid line, while
$\gamma$ dwells at the dotted line. The EOS of dark energy crosses
$-1$ at about $s=-0.22$, or $z=0.25$. {\bf{(b)}} The right panel:
the corresponding deceleration parameter, which crosses 0 at about
$s=-0.40$, or $z=0.49$. From \cite{self3}.}
 \label{rho2}
 \end{figure}

  One concludes from the above equation that a smaller $r_c$, a
 smaller $\Omega_{m0}$ (Recall that it is defined as the present value of
  the energy density of dust matter over the critical density), or
  a larger $\Omega_{ki}$ (which is defined as the present value of
  the kinetic energy density of the scalar  over the critical
  density) is helpful to shift the stagnation point to lower redshift
  region. We show a concrete numerical example of this crossing
  behaviours in  fig \ref{rho2}.
  For convenience we introduce the dimensionless density and rate of
  change with respect to redshift of dark energy as below,
 \bea
 \beta=\frac{\rho_{de}}{\rho_c}=\frac{\Omega_{r_c}}{b^2}\left[y_1^2+y_2^2+y_4^2
 -y_4^2(1+2\frac{y_1^2+y_2^2+y_3^2}{y_4^2})^{1/2}\right],
 \ena
 where $\rho_c$ denotes the present critical density of the universe, and
 \bea
 \nonumber
 \gamma &=&
 \frac{1}{\rho_c}\frac{y_4^2}{\Omega_{r_c}}\frac{d\rho_{de}}{ds}\\
 &=&3\left[(1+2\frac{y_1^2+y_2^2+y_3^2}{y_4^2})^{-1/2}(2y_1^2+y_3^2)-2y_1^2\right].
 \ena

 A significant parameters from the viewpoint of
  observations is the deceleration parameter $q$, which carries the total
  effects of cosmic fluids. We plot $q$ in these figures for corresponding density
   curve of dark energy. In the fig \ref{rho2} we set
   $\Omega_m=0.3$.
   $\Omega_{r_c}$ is defined as
   the present value of
  the energy density of $\rho_0$ over the critical density
  $\Omega_{r_c}={\rho_0}/{\rho_c}$.

 Now, we turn to the evolution of a universe with a phantom (\ref{lagphan}) in DGP.
 In an FRW universe the density and pressure of a phantom can be written
 as (\ref{rhophan}), (\ref{pphan}).

  To compare with the results of the ordinary scalar, here we set a
  same potential as before,
 \be
 U=U_0e^{-\lambda_2\frac{\psi}{\mu}}.
 \en

 The ratio of change of  density of virtual dark energy with respective to $\ln (1+z)$
  becomes,
 \bea
  \frac{d \rho_{de}}{d \ln
  (1+z)}= 3[-\dot{\psi}^2+\theta(1
  +\frac{-\dot{\psi}^2+2U+2\rho_{dm}}{\rho_0})
  ^{-1/2}(-\dot{\psi}^2+\rho_{dm})].
  \label{derivation ph}
  \ena
  To study the behaviour of the EOS of dark energy, we first take a
  look at the signs of the terms of RHS of the above equation.
  $(-\dot{\psi}^2+\rho_{dm})$ represents the total energy density of
  the cosmic fluids, which should be positive. The term $(1
  +\frac{-\dot{\psi}^2+2U+2\rho_{dm}}{\rho_0})
  ^{-1/2}$ should also be positive. Hence if $\theta=-1$,
   both terms of RHS are negative: it never goes
  to zero at finite time. Contrarily, if $\theta=1$, the two terms of RHS
  carry opposite sign: the EOS of dark
  energy is able to cross phantom divide. In the following of the present subsection
  we consider the branch of $\theta=1$.

 Now the Friedmann
 constraint becomes
 \be
 \label{constraint ph}
 -y_1^2+y_2^2+y_3^2+y_4^2+y_4^2\left(1+2\frac{-y_1^2+y_2^2+y_3^2}{y_4^2}\right)^{1/2}=1.
 \en
 Again, one will see that in
 reasonable regions of parameters, the EOS of dark energy crosses
 $-1$, but from below $-1$ to above $-1$.

\begin{figure}
\centering
 \includegraphics[totalheight=1.8in, angle=0]{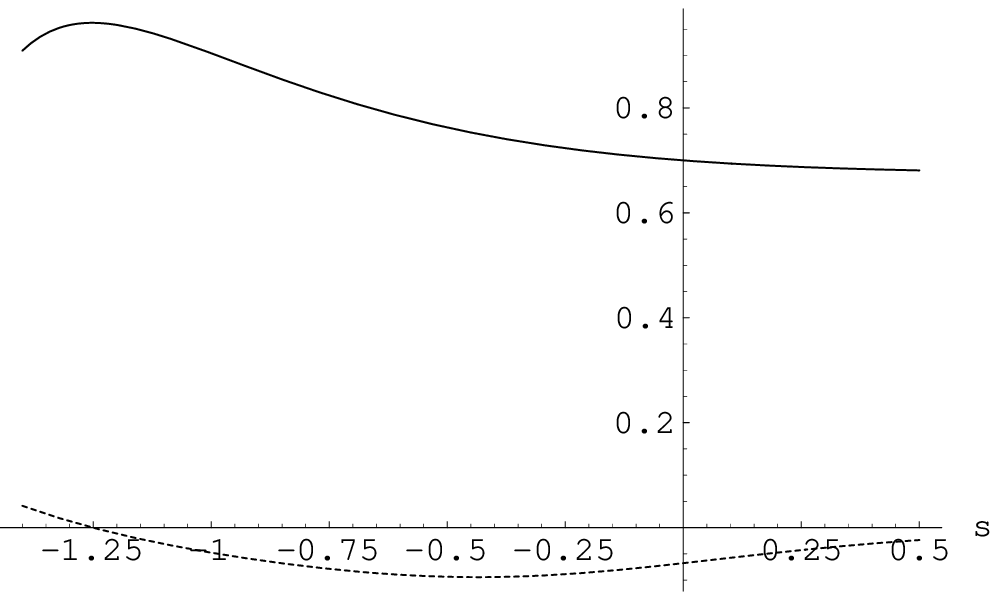}
\includegraphics[totalheight=2in, angle=0]{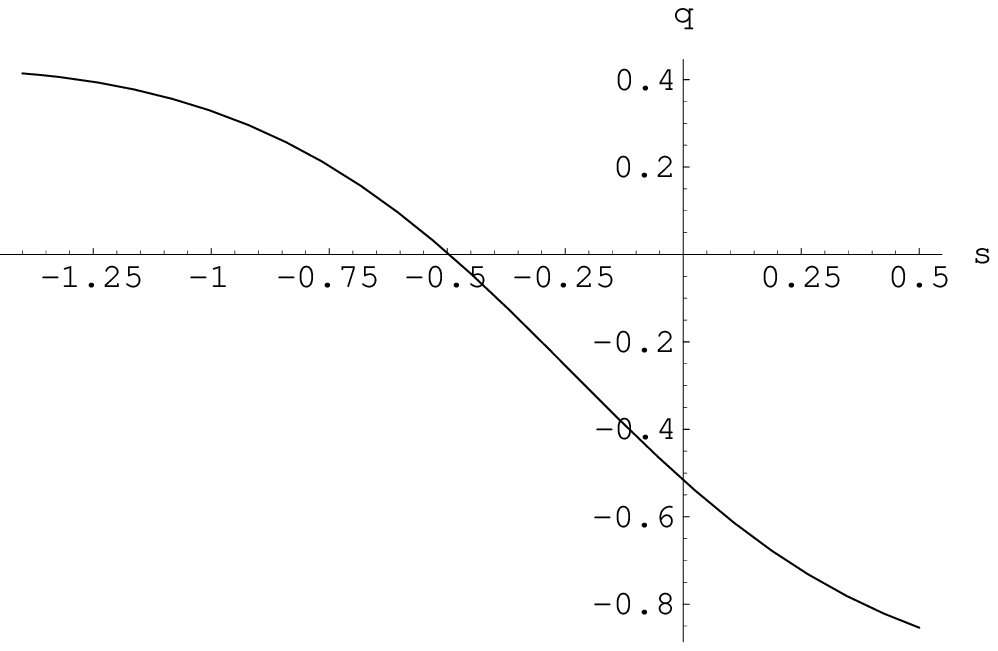}
\caption{For this figure, $\Omega_{ki}=0.01$, $\Omega_{r_c}=0.01$,
$\lambda=0.01$. {\bf{(a)}} The left panel: $\beta$ and $\gamma$ as
functions of $s$, in which $\beta$ resides on the solid line, while
$\gamma$ dwells at the dotted line. The EOS of dark energy crosses
$-1$ at about $s=-1.25$, or $z=1.49$. {\bf{(b)}} The right panel:
The corresponding deceleration parameter, which crosses 0 at about
$s=-0.50$, or $z=0.65$. From \cite{self3}.}
 \label{rho4}
 \end{figure}

 The stagnation point of $\rho_{de}$ inhabits at
 \be
 \frac{y_4}{\sqrt{2}-y_4}\left(-2+\frac{y_3^2}{y_1^2}\right)=2,
 \en
 which can be derived from (\ref{derivation ph}) and
 (\ref{constraint ph}). One concludes from the above equation that a smaller $r_c$, a
 smaller $\Omega_m$, or
  a larger $\Omega_{ki}$ is helpful to shift the stagnation point to lower redshift
  region, which is the same as the case of an ordinary scalar.
   Then we show a concrete numerical example of the crossing
  behaviour of this case in fig \ref{rho4}.
  The dimensionless density and rate of
  change with respect to redshift of dark energy become,
  \bea
 \beta=\frac{\rho_{de}}{\rho_c}=\frac{\Omega_{r_c}}{b^2}\left[-y_1^2+y_2^2+y_4^2
 +y_4^2(1+2\frac{-y_1^2+y_2^2+y_3^2}{y_4^2})^{1/2}\right],
 \ena
 and
 \be
 \gamma =3\left[-(1+2\frac{-y_1^2+y_2^2+y_3^2}{y_4^2})^{-1/2}(-2y_1^2+y_3^2)+2y_1^2\right].
 \en
  Similarly, the deceleration parameter
     is plotted in the figure \ref{rho4} for corresponding density
   curve of dark energy. In this figures we also set $\Omega_m=0.3$.

 Fig \ref{rho4} explicitly illuminates that the EOS of virtual dark energy
 crosses $-1$, as expected. At the same time the deceleration
 parameter is consistent with observations.

 \section{summary}

  The recent observations imply that the EOS of dark energy may cross $-1$.
  This is a remarkable phenomenon and attracts much theoretical attention.

  We review three typical models for the crossing behavior. They are
  two-field model, interacting model, and modified gravity model.

  There are several other interesting suggestions in or beyond the three
  categories mentioned above. We try to list them here for the
  future researches. We are apologized for this incomplete reference
  list on this topic.

 Almost in all dark energy models the dark energy  is suggested as scalar. However,
  3 orthogonal vectors can also play this role. For the interacting
  vector dark energy and phantom divide crossing, see
  \cite{weivector}.  For the suggestion of crossing the phantom divide with a
  spinor, see \cite{spin}. Multiple k-essence sources are helpful to fulfil the condition for
 phantom divide crossing \cite{multike}. Phantom divide  crossing can be realized by non-minimal coupling and Lorentz invariance
 violation \cite{lorevio}. An exact solution of a two-field model for this crossing has been
 found in \cite{exacttwo}.

   For previous interacting $X$ (quintessence or phantom) models with crossing $-1$, see \cite{internew}.
   The cosmology of interacting $X$  in loop gravity has been studied in
 \cite{Xloop}.

 Interacting holographic dark energy is a possible mechanism for the phantom divide
 crossing \cite{ihd}. And the thermodynamics of interacting holographic dark energy with phantom divide crossing
  is investigated in \cite{thermalihd}. An explicit model of $F(R)$ gravity in which the dark energy crosses
the phantom divide is reconstructed in \cite{f(R)}. The phantom-like
effects in a DGP-inspired  $F(R,\phi)$ gravity model is investigated
in \cite{phandgp}. Based on the recent progress in studies of source
of Taub space \cite{taubsource}, a new braneworld in the
sourced-Taub background is proposed \cite{taubbrane}, while the
previous  brane world models are imbedded in AdS (RS) or Minkowski
(DGP). In this model the EOS for the virtual dark energy of a dust
brane  in the source region can cross the
 phantom divide. For other suggestions in brane world model,
see \cite{branecorss}.

 Similar to the coincidence problem of dark energy, we can ask why
 the EOS crosses $-1$ recently? This problem is studied in \cite{wei2coin}.

 On the observational side, the present data only mildly favor the
 crossing behavior. We need more data to confirm or exclude it.

 Theoretically, we should find more natural model which has less
 parameters. We must go beyond the standard model of particle
 physics. The problem cosmic acceleration is a pivotal problem to access new physics. To study the problem
 of crossing $-1$ EOS will impel the investigation to the new Laws of nature.

  {\bf Acknowledgments}
  We thank to all the original authors who permit us to use their figures.  After submission of this invited review,
  Our special thankfulness goes to Y. Cai, who informed me that their review on quintom cosmology would appear soon. And Y Cai give us several beneficial suggestions.
  Hence this article is slightly different from the published version.

\end{document}